\documentclass[review,3p]{elsarticle}

\usepackage{amssymb}
\usepackage{amsmath}
\usepackage[utf8]{inputenc}
\usepackage[ruled,vlined,linesnumbered]{algorithm2e}   
\usepackage{listings}                           
\usepackage{newtxmath}                          
\usepackage{mathdots}
\usepackage{mathtools}

\usepackage{enumitem}
\usepackage{tikz}
\usepackage{arydshln}
\usepackage{circuitikz}
\usepackage{orcidlink}
\usetikzlibrary{arrows}
\usetikzlibrary{calc}
\usetikzlibrary{intersections}
\usetikzlibrary{patterns}
\usepackage{contour}
\usepackage{wrapfig}
\usepackage[T1]{fontenc}
\usepackage{subcaption}
\definecolor{darkgreen}{RGB}{34, 139, 34}
\definecolor{darkblue}{RGB}{116, 116, 200}
\definecolor{darkblue1}{RGB}{56,58,244}
\definecolor{darkred}{RGB}{253, 87, 87}
\definecolor{darkred1}{RGB}{238,53,57}
\definecolor{lightblue}{RGB}{0, 138, 255}
\definecolor{darkyellow}{RGB}{237,202,22}

\usepackage{overpic}
\definecolor{clrAG}{HTML}{7B3F00}                      

\usepackage{soul}
\usepackage{lineno}

\newcommand{\bluer}[1]{
\marginpar{\color{blue} $\Longleftarrow$}
\begingroup\color{blue}#1\endgroup}
\newcommand{\bluevar}[1]{
\begingroup\color{blue}#1\endgroup}
\renewcommand{\bluevar}[1]{#1}
\renewcommand{\bluer}[1]{#1}

\journal{Computer Aided Geometric Design}

\begin{document}
\fntext[eq]{Joint first authors.}
\begin{frontmatter}

\title{
Unlocking Euclidean Problems with Isotropic Initialization}


\author[kaust]{\corref{cor1}\fnref{eq} Khusrav Yorov\orcidlink{0009-0008-6296-8718}}
\ead{khusrav.yorov@kaust.edu.sa}
\author[casa,rwth]{\fnref{eq} Bolun Wang\orcidlink{0000-0002-2027-870X}}
\ead{bolun.wang@rwth-aachen.de}
\author[kaust]{Mikhail Skopenkov\orcidlink{0000-0003-2453-0009}}
\ead{mikhail.skopenkov@gmail.com}
\author[tuw]{Helmut Pottmann\orcidlink{0000-0002-3195-9316}}
\ead{helmut.pottmann@gmail.com}
\author[xjtu]{Caigui Jiang\orcidlink{0000-0002-1342-4094}}
\ead{cgjiang@xjtu.edu.cn}
\address[kaust]{King Abdullah University of Science and Technology (KAUST), Thuwal 23955, Saudi Arabia}
\address[casa]{MAIS, Institute of Automation, Chinese Academy of Sciences, Beijing 100190, China}
\address[rwth]{RWTH Aachen University, 52062 Aachen, Germany}
\address[tuw]{TU Wien, 1040 Vienna, Austria}
\address[xjtu]{Xi'an Jiaotong University, Xi'an, China}
\cortext[cor1]{Corresponding author.}

\begin{abstract}
Many problems in Euclidean geometry, arising in computational design and fabrication, amount to a system of constraints, which is challenging to solve. We suggest a new general approach to the solution, which is to start with analogous problems in isotropic geometry. Isotropic geometry can be viewed as a structure-preserving simplification of Euclidean geometry. The solutions found in the isotropic case give insight and can initialize optimization algorithms to solve the original Euclidean problems.  
We illustrate this general approach with three examples: quad-mesh mechanisms, composite asymptotic-geodesic gridshells, and asymptotic gridshells with constant node angle.
\end{abstract}





\end{frontmatter}



\section{Introduction\label{sec:introduction}}

\begin{figure}[tb]
    \centering
    \begin{overpic}[width=0.92\textwidth]{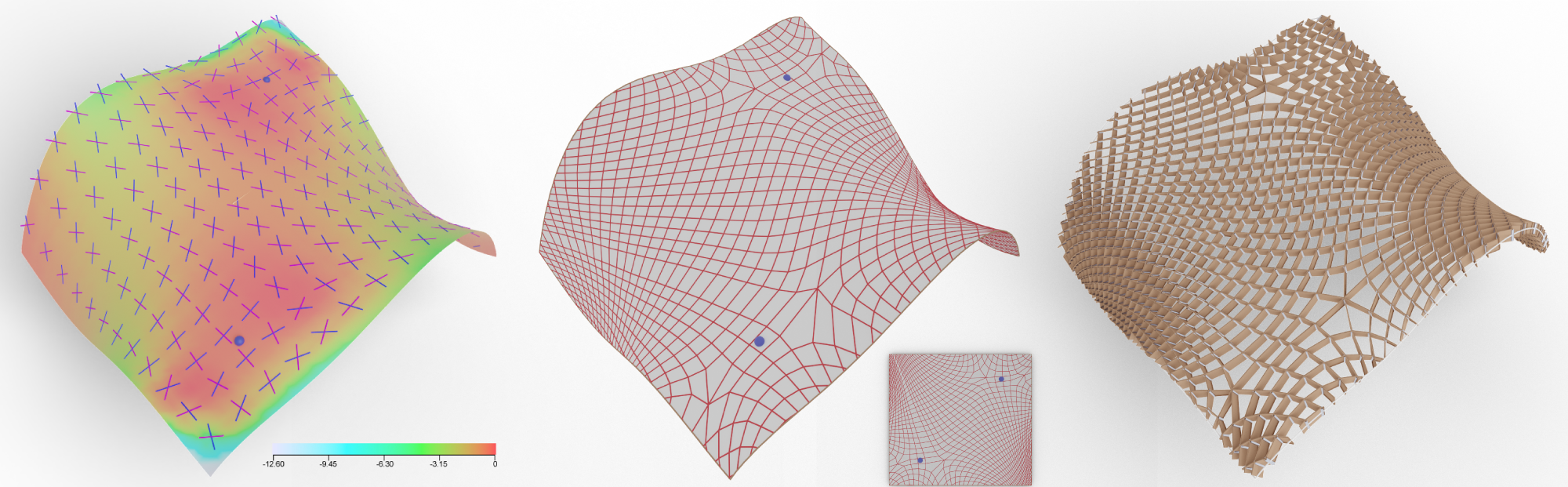} 
    \small
\put(17,3.2){\contour{white}{\tiny Gaussian curvature $\kappa$}}
\put(88,0){\includegraphics[width=1.9cm]{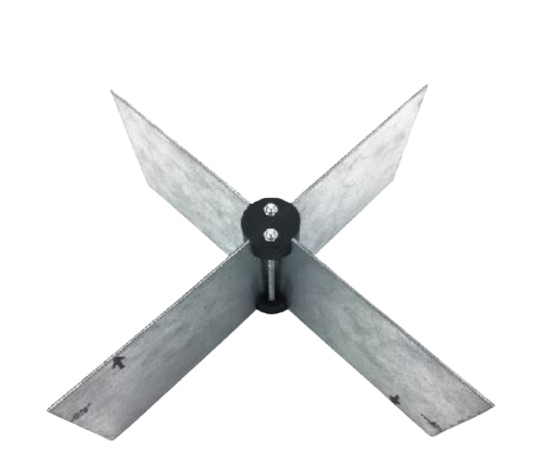}}
    \end{overpic}
    \captionsetup{width=0.92\linewidth}
    \caption{Design of a so-called asymptotic gridshell 
    with prescribed constant node angle and 
    positions of combinatorial singularities by isotropic initialization.  An \emph{asymptotic gridshell} (right) is obtained by bending originally straight lamellas and arranging them in a quadrilateral grid so that they are orthogonal to a reference surface. This forces the lamellas to follow the asymptotic directions of the surface. 
    A constant intersection angle of lamellas 
    simplifies fabrication, \bluevar{as it allows the use of identical nodes throughout the structure (inset)
    }. 
    Left: A surface with a nearly constant \emph{isotropic} angle between asymptotic directions (
    crosses) and flat points near prescribed positions (
    blue dots). Color shows the variation of Euclidean Gaussian curvature. Such a surface is given by a simple analytic expression in isotropic geometry, but not in Euclidean geometry. 
    Middle: The remeshed surface and its top view (
    in the inset). Singularities appear around the blue points where the flat points are prescribed.
    Right: An asymptotic gridshell with a constant \emph{Euclidean} angle \(72^\circ\) between the lamellas obtained by optimization initialized by the mesh in the middle. 
    %
    }
    \label{fig:teaser}
\end{figure}

\begin{figure}[htbp]
    \centering
\begin{overpic}
[width=1.0\linewidth]{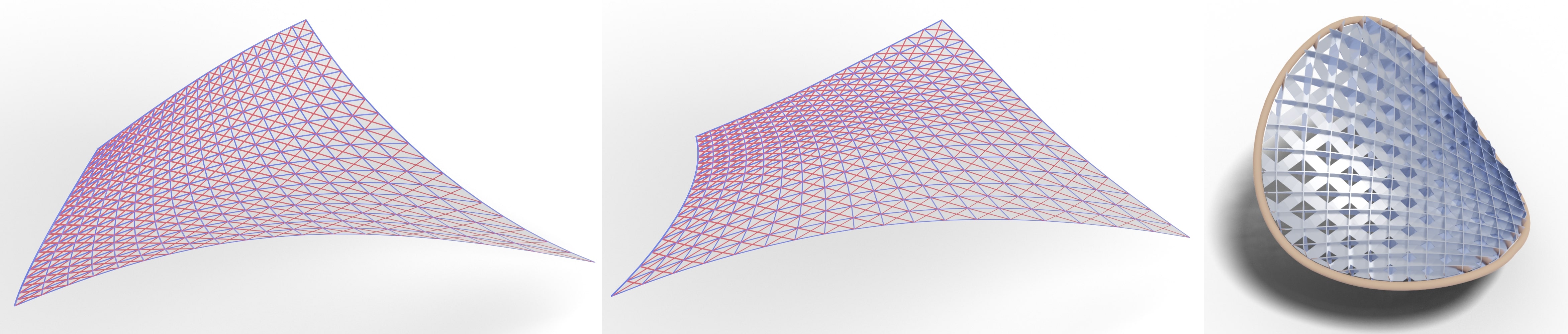}
\small
\put(0,17){\contour{white}{(a)}}
\put(40,17){\contour{white}{(b)}}
\put(77,17){\contour{white}{(c)}}
 \end{overpic}
\caption{Design of a so-called \emph{composite asymptotic-geodesic gridshell} by isotropic initialization. In particular, \emph{an AGAG gridshell} (c) is obtained by bending originally flat lamellas and arranging them so that some are orthogonal and some are tangent to a reference surface, and four lamellas meet at each node. This forces the lamellas to follow the asymptotic (red) and geodesic (blue) curves on the surface respectively, organized in a so-called AGAG web. We construct an isotropic AGAG web (a) and then optimize it to a Euclidean one (b).
The meshes were downsampled by removing some polylines from the red and blue families to reduce density while keeping their overall shape. 
Cutting 
along a closed curve 
leads to a gridshell 
(c). Isotropic AGAG webs are completely classified, 
but it is unknown if a single nontrivial smooth Euclidean AGAG web exists. 
}
\label{fig:AGAG1}
\end{figure}

Many problems in computational design and fabrication amount to the solution
of a system of constraints which is solved by numerical optimization. Despite
the presence of constraints, the solution space or design space may still
be sufficiently large to be explored by a designer. However, it may be difficult to access this design space. This is equivalent to finding
an initial guess for the underlying optimization problem. Especially
difficult cases are those where one has very few, or 
even no explicit solution. An example for the latter situation is provided by
hybrid asymptotic-geodesic gridshells. See Figs.~\ref{fig:teaser}--\ref{fig:AGAG1}.

There are various ways to handle such difficult situations. One
may initially omit some constraints, possibly find a solution then,
and use that to initialize optimization. One may also initialize with a nearly
trivial special case, with the drawback that
this may not open a path towards interesting solutions.

In the present paper, we suggest the following general two-step approach to certain hard problems in Euclidean geometry, allowing us to overcome this obstacle. 

In the first step, we solve the same problem in isotropic geometry. Isotropic geometry is one of the classical non-Euclidean geometries, 
and can be viewed as a structure-preserving simplification of Euclidean geometry.
It is obtained by replacing the Euclidean norm $\sqrt{x^2+y^2+z^2}$ in space with the simpler isotropic semi-norm~$\sqrt{x^2+y^2}$. 
That alone would be too degenerate. However, isotropic geometry is also based on a
6-dimensional group of isotropic congruence transformations, which
leads to a wealth of results that are very similar to their Euclidean
counterparts  \cite{Sachs:1990}. 

The second step is to use the isotropic solution to initialize an optimization algorithm for a numerical solution to the original Euclidean problem. 
Last but not least, the isotropic solution may provide insight into how the Euclidean problem can be solved.

This approach can be traced back to classical works. An example is the Plateau problem on finding 
a minimal surface with a given boundary. 
This problem was solved in a fairly general setting by \bluer{M\"untz}~\cite{Muntz}. 
He constructed minimal surfaces by deformation of the graphs of harmonic functions. The latter are minimal surfaces in isotropic geometry. So, in this case, the optimization led to the whole existence proof and a construction method.
\begin{figure}[tb]
    \centering
\begin{overpic}
[width=0.92\linewidth]{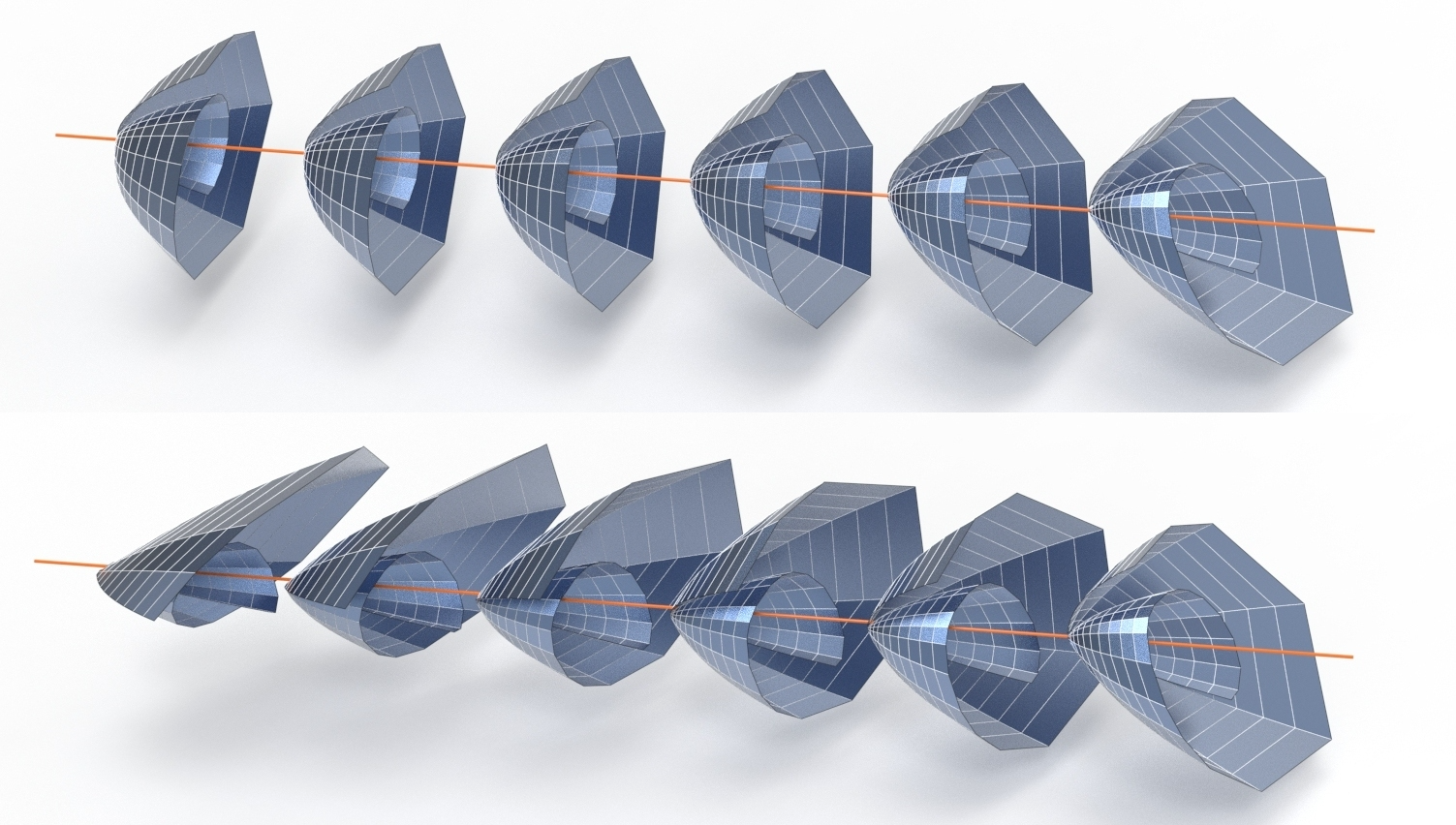}
\small
\put(0,25){\contour{white}{(b)}}
\put(0,52){\contour{white}{(a)}}
 \end{overpic}
 \captionsetup{width=0.92\linewidth}
\caption{
Design of mechanisms with rigid faces and rotational joints in edges by isotropic initialization. We construct an isotropic mechanism (a) and then optimize it to a Euclidean one (b). 
A few positions of each mechanism are shown. The line depicts the isotropic direction. 
While the initial (rightmost) positions of 
two mechanisms 
are visually the same, 
the flexion leads to 
different shapes. 
Isotropic mechanisms are completely classified, whereas the construction of Euclidean ones is a widely open problem. 
The particular mechanism in (a) belongs to class (i) introduced in Section~\ref{sec:flexible}.   
}
\label{fig:flexible-isotropic-i}
\end{figure}
The suggested approach works surprisingly well for various practical geometric problems. 

We illustrate it with examples, which are nearly inaccessible by the methods available before: construction of quad-mesh mechanisms (see Fig.~\ref{fig:flexible-isotropic-i}), composite asymptotic-geodesic gridshells (see Fig.~\ref{fig:AGAG1}), and asymptotic gridshells with constant node angle (see Fig.~\ref{fig:teaser} and Fig.~\ref{fig:CRPC1}).

The presented method is not a general-purpose simple trick to solve
all kinds of hard problems in Euclidean geometry. It naturally has certain
limitations. Most importantly, one must be able to come up with a meaningful isotropic initial
guess
that is not too degenerated to be transformed into a Euclidean solution. Our 
examples are interesting in various ways. This may not be straightforward, as 
is illustrated in some of our examples, especially on AGAG webs and quad mesh
mechanisms. The latter required a new non-degenerate definition of
isometric surfaces in isotropic 3-space \cite{isometric-isotropic}. 




\subsection{Related work}

Fundamental contributions to isotropic geometry \cite{strubecker1,strubecker2,strubecker3}
and the monograph \cite{Sachs:1990} are written in German. Thus, we point to further contributions
that contain a summary of key concepts, namely \cite{pottmann1994,pottmann-2007-ds,pottmann-2009-lms,BMW-2024}. 

In the following, we first concentrate on known applications of isotropic geometry and then turn to prior work on the applications treated in this paper. For those,
the use of isotropic geometry is new. 

\smallskip
\noindent \emph{Structural design.} K.~Strubecker \cite{Strubecker1962} pointed out that the so-called Airy stress function, which is associated with a planar equilibrium state, is best understood when viewing its graph  within isotropic geometry. There is a complete correspondence between mechanical properties, e.g. stresses, and properties of the Airy stress surface in isotropic geometry, e.g. isotropic curvatures. 
Following up on that, various contributions to structural design, in particular
when considering only vertical loads, exploited the geometry of the stress surface for
design \cite{Chiang2022_DD,Chiang2022_IJSS,Miki2015,Millar2022,Tellier2021,vouga-2012-sss}. With constant vertical loads, stress surface and design surface may
agree up to vertical scaling \cite{Millar2022} and then 
represent surfaces of constant isotropic Gaussian curvature, for which elegant geometric constructions are known
\cite{strubecker2,strubecker4,strubecker5}. These self-Airy surfaces provide
many design options for the construction of polyhedral surfaces in equilibrium
with forces only in edges, as needed for steel-glass structures \cite{Millar2022}. 
Even if such approaches are based on very
rough approximations of real loads, these designs can later be changed via
optimization towards more realistic load cases. Such an approach is in the spirit of the present paper. 

Beyond 
applications in connection
with the Airy stress function, isotropic geometry has been employed 
for geometric approaches to functions
defined by 2D images \cite{koenderink}
or functions defined on surfaces
\cite{pottmann1994}.

\smallskip
\noindent \emph{Visual appearance of polyhedral surfaces.} 
Static equilibrium of a planar truss is related to a stress
surface with planar faces. Stress in the members is
equivalent to the isotropic dihedral angles of the polyhedral
stress surface. This implies that material-minimizing trusses have
 the smoothest polyhedral surfaces in isotropic space 
as stress surfaces \cite{pellis-optimal-2017}. They are
 discrete versions of isotropic  principal parameterizations.
 Motivated by this result, which follows naturally from the pioneering
 work of \cite{michell-1904}, Pellis et al. \cite{pellis-smooth-2019} studied the
 visual smoothness of polyhedral surfaces in Euclidean geometry.
 With an appropriate smoothness measure, principal meshes yield
 highest visual smoothness. As in the isotropic
 case, surfaces of minimal total absolute curvature $\int (|\kappa_1| +|\kappa_2|)dA$
 ($\kappa_i$ being principal curvatures) yield the optimal shapes in 
 terms of visual smoothness (
 see \cite{pellis-thesis}). Isotropic
 geometry also played a key role in the design of polyhedral surfaces
 with controlled roughness \cite{BMW-2024}.

\smallskip
\noindent \emph{Laguerre sphere geometry.}  Laguerre
geometry is one of the classical sphere geometries. It
comes with several models, one of which plays in 
(appropriately extended) isotropic 3-space. There it
appears as isotropic counterpart of Euclidean M\"obius
geometry. For example, a conical mesh is an object
of Laguerre geometry and appears in the isotropic model
as (isotropic) circular mesh \cite{pottmann-2007-ds}. 
Some problems, such as the
study of rational curves and surfaces with rational
offsets, are greatly simplified when using the isotropic
model of Laguerre geometry \cite{peternell-1998-lgaro}. 
Laguerre minimal surfaces become more easily accessible in
the isotropic model, where they appear as graphs of
biharmonic functions  \cite{pottmann-2009-lms, skopenkov-2012-ruledlag}. Envelopes
of moving cones or cylinders are well treated in Laguerre
geometry and the isotropic model, leading to an
application in CNC machining \cite{CNC-skopenkov-2020}.

We now turn to a short outline of prior work on those applications
which we later introduce to demonstrate the use of isotropic geometry for
the initialization of difficult Euclidean optimization problems.

\smallskip
\noindent \emph{Quad mesh mechanisms.} Studies of flexible polyhedral
surfaces have a long history, and recently received growing interest as
examples of transformable designs. We address here quad meshes, mostly
with planar faces, which are mechanisms with rigid faces and rotational
joints in the edges. The problem is algebraic in nature, but difficult.
So far, only
mechanisms with $3 \times 3$ planar faces have been classified  \cite{izmestiev-2007}, \bluer{and understanding of individual classes is still in progress \cite{Nurmatov-etal-26}.} 
The composition to larger meshes is largely
unclear, except for very special cases (see e.g. \cite{sauer:1970,he2020rigid, Nawratil25}).
We develop a recent approach based on optimization  \cite{quadmech-2024}, which contains a much more complete overview of prior
work and a variety of strategies for
initialization. In the present paper, we propose another such
strategy; see Fig.~\ref{fig:flexible-isotropic-i} and  Section~\ref{sec:flexible}.

\smallskip
\noindent \emph{Gridshells from straight lamellas.} Eike Schling and his
coworkers designed and fabricated gridshells by bending straight flat
lamellas and arranging them in a quadrilateral grid such that the lamellas
are orthogonal to an underlying reference surface $S$ \cite{schling:2018,schling2022geometry}. Thus, in their
final position the lamellas follow asymptotic curves on $S$. If lamellas
meet at right node angles, $S$ is a minimal
surface. For another constant node angle $\gamma$, the surface $S$ is a surface with constant ratio of principal curvatures  $\kappa_1/\kappa_2=\tan (\gamma/2) <0$ (\emph{CRPC surface}). 
This is equivalent to the constant ratio $H^2/K=-\cot^2\gamma$ of squared mean curvature and Gaussian curvature. 
One of the key advantages of such a gridshell is that they can be actuated to realize highly controlled transformable structures~\cite{schikore2021kinetics}. Only very special cases of such natural generalizations of minimal surfaces are explicitly known \cite{helCRPC-2022}, but
there is work on the computational design via optimization \cite{CRPC-2021,quad-aag-2020}. 
We propose another strategy; see Fig.~\ref{fig:CRPC1}.

Many more open questions arise if one uses three or four families of lamellas and
arranges them in form of a 3-web or 4-web, respectively. 
For such
structures, some families of 
lamellas are tangential to $S$ and thus follow geodesics on $S$. This leads to various types of these asymptotic-geodesic webs. Computational models based on discrete differential geometry have been proposed in \cite{Eike-CAD-2022,rect-strip-2023, web-propagation-2025}, but an understanding
of potential shapes and the initialization of optimization is still a 
challenge. We are not aware of any explicit
example of a surface which carries a non-trivial 4-web of type AGAG (see Figs. \ref{fig:AGAG1}(b) and  \ref{fig:AGAG_merged}(b, f)),
making even the computation via optimization a challenge due to a lack of
reasonable initial guesses. 
Since discrete AGAG-webs have been classified in isotropic geometry  \cite{AGAG-2024}, they will be used for that purpose (Section~\ref{sec:webs}).

\subsection{Overview and contributions}
We discuss the initialization of optimization algorithms for the numerical solution of Euclidean
geometric problems based on solutions of the isotropic counterparts. While the latter are expected to
be easier to deal with, the solutions in isotropic
geometry are not straightforward in the cases
which we outline in the present paper.

After a short summary of some essential background
from isotropic geometry (Section~\ref{sec:preliminaries}),
we turn to quad mesh mechanisms in Section~\ref{sec:flexible}. Very
recent work by \cite{isometric-isotropic} on a fruitful definition of isometric maps between surfaces
in isotropic geometry allows one to define and fully characterize
flexible quad meshes with planar faces in isotropic geometry \cite{pirahmad2025}.
The resulting isotropic mechanisms are already surprisingly close to the
Euclidean ones and provide a good basis for computational design, also because
many isotropic mechanisms fall into a class of meshes that has been studied
previously for architectural applications \cite{PP-2022}.

In Section~\ref{sec:webs}, we turn to gridshells that are formed by bending
straight flat lamellas and geometrically require the computation of webs
from geodesic and asymptotic curves \cite{Eike-CAD-2022}. 3-webs of geodesics can be easily computed in isotropic
geometry based on a theorem by Graf and Sauer \cite{graf:sauer:1924}. 

However, others are harder to deal with. 
We construct isotropic counterparts of 3-webs formed
by two families of asymptotic and one family of geodesics. 
Fortunately, for AGAG webs, i.e., 
 4-webs formed by two geodesic and two asymptotic curve families, 
 \begin{figure}[t]
    \centering
 \begin{overpic}
[width=1.0\linewidth]{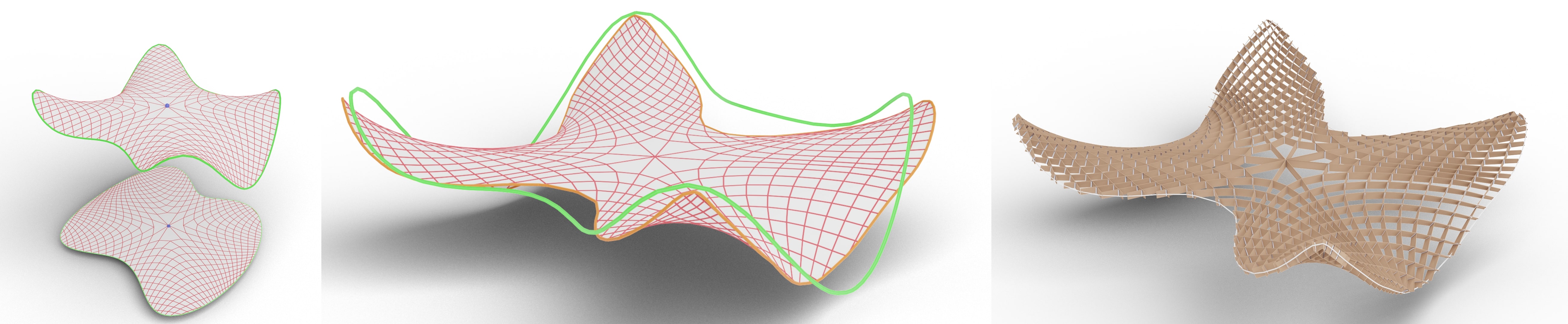}
\small
\put(0,20){\contour{white}{(a)}}
\put(21,20){\contour{white}{(b)}}
\put(62,20){\contour{white}{(c)}}
 \end{overpic}
\caption{Design of an asymptotic gridshell with prescribed constant node angle ($60^\circ$), 
boundary curve (green), and 
flat point (blue dot) by isotropic initialization.
(a) The approximate isotropic CRPC surface constructed using Algorithm~\ref{alg:boundary+features} and its top view. 
(b) The Euclidean CRPC surface with a constant angle \(60^\circ\) between the asymptotic polylines obtained by optimization initialized by the mesh in (a).
(c) An asymptotic gridshell extracted from (b). }
\label{fig:CRPC1}
\end{figure}
we can use the recent complete classification of
their isotropic counterparts \cite{AGAG-2024}. 
The AGAG case is the most challenging. To our knowledge, no smooth Euclidean AGAG webs are known. \bluer{As for discrete AGAG webs,} only a few numerical examples are shown by Schling et al.~\cite{Eike-CAD-2022}; however, those examples are mostly flat and symmetric due to the lack of proper initialization. 
By our approach, we obtain complex Euclidean AGAG shapes (see Figs.~\ref{fig:AGAG1}, \ref{fig:AGAG_merged}, \ref{fig:AGAG2grid}). 

Our final application deals with the computation of asymptotic gridshells with
constant node angle, in particular those that approximate a given boundary curve;
see Fig.~\ref{fig:CRPC1} and Section~\ref{sec:CRPC}.
Geometrically, this amounts to the computation of a surface with a constant ratio
of principal curvatures. In isotropic geometry, the problem is simplified
by a novel approximation of such surfaces using complex analysis. 
Using this setup, we construct these surfaces with given flat points, either with a given boundary or without (see Figs.~\ref{fig:teaser}, ~\ref{fig:CRPC1},  ~\ref{fig:CRPC3},~\ref{fig:CRPC2}). Here flat points are those points where more than two asymptotic lines can meet.

Details on the optimization are provided in Section~\ref{sec:optimization}
and their performance, choice of parameters, and a discussion of results
and future work form the content of Section~\ref{sec:conclusion}.

\section{Preliminaries\label{sec:preliminaries}}



\subsection{Isotropic geometry}

In \emph{isotropic geometry}, the Euclidean norm $\|(x,y,z)\|=\sqrt{x^2+y^2+z^2}$ in space with the coordinates $x,y,z$ is replaced with the \emph{isotropic semi-norm} $\|(x,y,z)\|_i:=\sqrt{x^2+y^2}$. 
The \emph{isotropic congruence transformations} are special affine transformations preserving the isotropic semi-norm:
%
\begin{equation*}
  \mathbf{x}'= A\cdot \mathbf{x}+\mathbf{b}, \quad 
A=  \begin{pmatrix} \cos\phi & -\sin\phi & 0  \\
 \sin\phi & \cos\phi & 0  \\
 c_1 & c_2 & 1 \end{pmatrix} 
\end{equation*}
for some values of the parameters $\mathbf{b}\in\mathbb{R}^3$ and $\phi,c_1,c_2\in\mathbb{R}$. 

These transformations appear as Euclidean congruences in the projection onto the plane $z=0$, which we call \emph{top view}. The top view of a point $P$ 
is denoted by $\overline P$. 
For instance, two planar quadrilaterals with the same top view (but not in one isotropic plane) are isotropically congruent. 
\emph{Isotropic distances} between
points, \emph{isotropic angles} between 
lines, and \emph{isotropic areas} of polygons are expressed through the isotropic norm in the usual way and appear in the top view as Euclidean distances, angles, and areas. 
Points $P$ and $Q$ with the same top view are called \emph{parallel}. The isotropic distance $\|\overrightarrow{PQ}\|_i$ between them vanishes. Lines and planes parallel to the $z$-axis are called \emph{isotropic}. See \cite{petrov-tikhomirov-06} for an elementary introduction.


One cannot define the isotropic angle between (non-isotropic) planes in the same way because their top views are just the same. Whenever a direct analog of Euclidean definition degenerates, one needs to find a replacing, nondegenerate one. Here the right definition of the \emph{isotropic angle} is the difference of the slopes of the two planes. An isotropic plane is viewed as \emph{orthogonal} to any non-isotorpic plane.

An isotropic sphere is the set of points at a fixed isotropic distance from a given point. It looks like a cylinder, hence called \emph{isotropic sphere of cylindrical type}. There is another type: the surface of constant normal curvature $A\ne 0$, given by the  equation $2z = A(x^2 + y^2)+Bx+Cz+D$ for some $B ,C ,D$. In a Euclidean interpretation, it is 
a paraboloid of revolution, hence called \emph{isotropic sphere of parabolic type}.

In contrast to Euclidean geometry, isotropic geometry possesses a metric duality. It is defined as the polarity with respect to the \emph{unit isotropic sphere} $2z = x^2 + y^2$, which maps a point 
$(x^*,y^*,z^*)$
to the plane 
with the equation $z + z^*= xx^*+yy^*$, 
and vice versa. 
The point dual to a non-isotropic plane $p$ is denoted by $p^*$, and vice versa. 
The isotropic angle between two non-isotropic planes equals the isotropic distance between the points dual to them. 

Isotropic lines and planes play a special role and are usually excluded as tangent spaces in differential geometry. 
A surface is \emph{admissible} if all its tangent planes are non-isotropic. 
An admissible surface can be locally represented as the graph of a function
$ z=f(x,y). $ 

Since isotropic distances are measured in the top view, the shortest paths on the surface, 
\emph{isotropic geodesics}, have straight top views. Thus isotropic geodesics are sections of the surface by isotropic planes. 

The \emph{isotropic Gauss map} takes a point on the surface to the point on the isotropic unit sphere such that the tangent planes at the two points are parallel. If this map is 1--1, then the oriented isotropic area of the image is called the \emph{total isotropic Gaussian curvature} of the surface. The latter area equals the oriented isotropic area of the \emph{metric dual surface} consisting of the points metric dual to the tangent planes of the original surface. The total curvature can also be computed as $\int K\, dxdy$, where $K:=f_{xx}f_{yy}-f_{xy}^2$ is the \emph{isotropic Gaussian curvature} of the surface $z=f(x,y)$. 


\subsection{Discrete differential geometry}


We also use \emph{discrete differential geometry} studying discrete analogs of curves and surfaces \cite{bobenko-2008-ddg}. Now we introduce basic notions and elaborate on them in subsequent sections.

By a \emph{discrete curve} we mean a collection of $m+1$ distinct points $f_{i}\in \mathbb{R}^3$ indexed by an integer $0\le i\le m$. A discrete curve determines a polyline with the vertices $f_{i}$ and edges (sides) $f_{i}f_{i+1}$.

By an $m\times n$ \emph{net} (\emph{discrete surface}) we mean a collection of $(m+1)(n+1)$ distinct points $f_{ij}\in \mathbb{R}^3$ indexed by two integers $0\le i\le m$ and $0\le j\le n$.
An $m\times n$ net determines a quad mesh with regular combinatorics: points $f_{ij}$ are \emph{vertices}, segments of the form $f_{ij}f_{i+1,j}$ and $f_{ij}f_{i,j+1}$ are \emph{edges},
and polylines of the form $f_{ij}f_{i+1,j}f_{i+1,j+1}f_{i,j+1}$ are \emph{faces}.
A vertex $f_{ij}$ with $i\in\{0,m\}$ or $j\in\{0,n\}$ is a \emph{boundary} vertex.
The polylines $f_{i0}\dots f_{in}$ for $0\le i\le m$ and $f_{0j}\dots f_{mj}$ for $0\le j\le n$ are called \emph{discrete parameter lines}  (namely, $i$-\emph{lines} and $j$-\emph{lines} respectively).

\begin{figure}[t]
    \centering
 \begin{overpic}[width=0.92\linewidth]{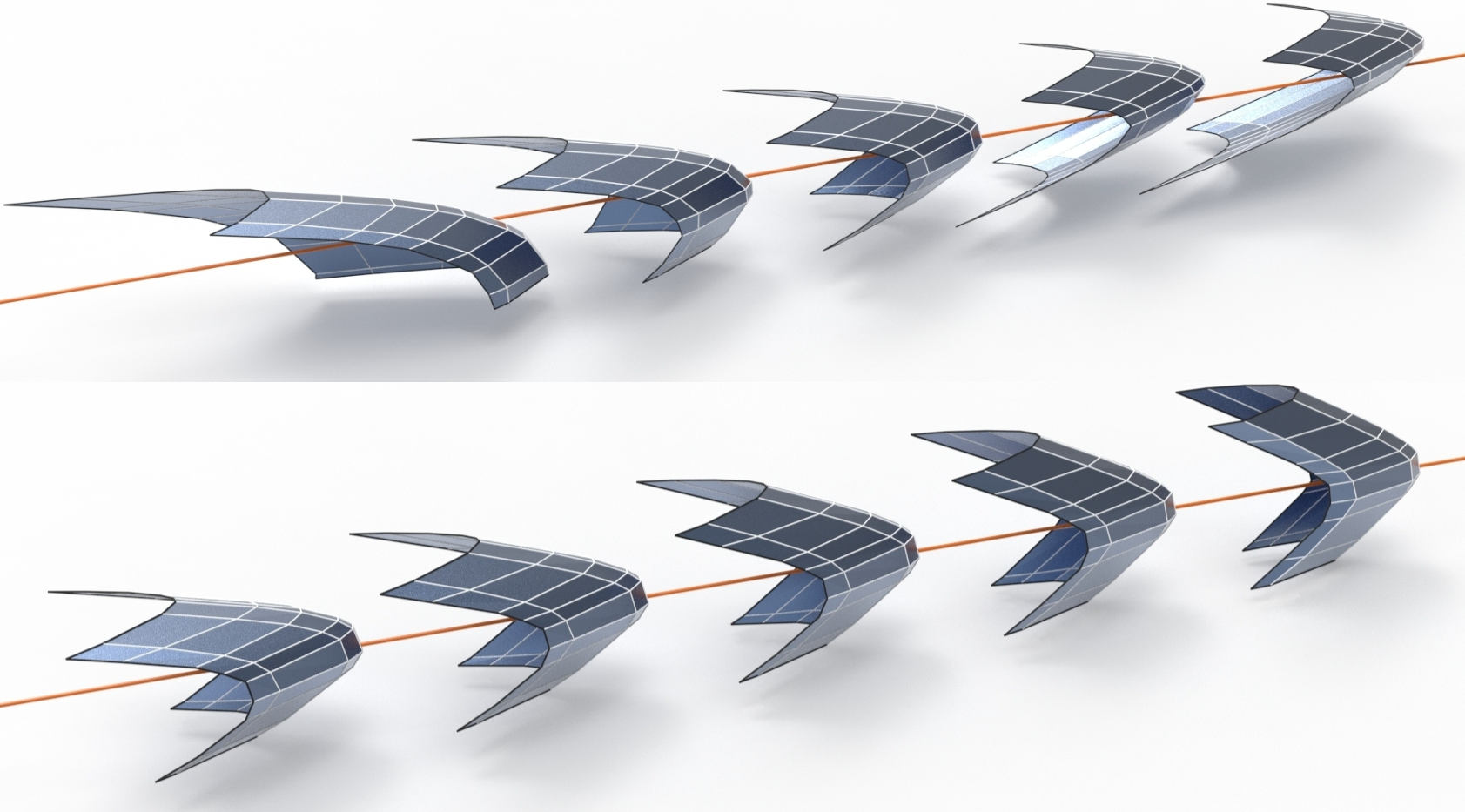}
 \small
\put(0,53){\contour{white}{(a)}}
\small
\put(0,25){\contour{white}{(b)}}
\end{overpic}
\captionsetup{width=0.97\linewidth}
\caption{(a) A Q-net that is flexible in isotropic geometry. A few positions of the isotropic mechanism are shown. The line depicts the isotropic direction. The net belongs to class (ii) described in Section~\ref{sec:flexible}. 
A single position of the isotropic mechanism (a) is used to initialize optimization, leading to the Euclidean mechanism (b). Although the optimization does not change the shape of this position much, the flexion leads to different shapes.
}
\label{fig:flexible-isotropic-ii}
\end{figure}

\indent In general, faces do not need to be planar. If all the faces are planar (but none is contained in a line), then the $m\times n$ net is called a \emph{Q-net}, and the faces are referred to as \emph{quadrilaterals} (although they can have self-intersections). See Fig.~\ref{fig:flexible-isotropic-ii}.

\section{Flexible polyhedral nets} \label{sec:flexible}

The first application of our method is the construction of quad-mesh mechanisms, also known as 
flexible nets. We are mostly interested in flexible polyhedral 
nets, i.e. flexible Q-nets.
We build on the classification of flexible Q-nets in isotropic geometry from \cite{pirahmad2025} (see Figs.~\ref{fig:flexible-isotropic-i} and~\ref{fig:flexible-isotropic-ii}). Let us briefly state 
this classification. 

First, we need to define what \emph{is} a flexible net in isotropic geometry. Even this step is non-trivial and has only recently been done \cite{isometric-isotropic}.  Only requiring rigid faces (\emph{face condition}) and rotational joints in the edges is not enough in isotropic geometry; otherwise, \emph{any} net would be flexible. 
The insight 
comes from the famous Gauss Theorema Egregium, saying that flexion of smooth surfaces preserves the Gaussian curvature. So it is natural to additionally require preservation of 
curvature at the vertices (\emph{vertex condition}).

This curvature is defined as follows. 
For an $m\times n$ Q-net $f_{ij}$, denote by $p_{ij}$ the plane of the face $f_{ij}f_{i+1,j}f_{i+1,j+1}f_{i,j+1}$. Consider the planes $p_1:=p_{i-1,j-1},p_2:=p_{i,j-1},p_3:=p_{ij},p_4:=p_{i-1,j}$ of consecutive faces around a particular non-boundary vertex $f_{ij}$ (see Fig.~\ref{figure:isotopic-gaussian-curvature}(a)). The top views of the points metric dual to those planes 
form a quadrilateral 
$\overline{p_1^*}\,\overline{p_2^*}\,\overline{p_3^*}\,\overline{p_4^*}$ called the \emph{top view of the 
face metric dual to the vertex $f_{ij}$} (see Fig.~\ref{figure:isotopic-gaussian-curvature}(b)). The oriented area of this quadrilateral is called the \emph{discrete isotropic Gaussian curvature}, 
or simply the \emph{curvature at} $f_{ij}$:
\begin{equation}\label{eq-def-curvature}
\Omega(f_{ij}):= \mathrm{Area}\left(\overline{p_1^*}\,\overline{p_2^*}\,\overline{p_3^*}\,\overline{p_4^*}\right)
=\frac{1}{2} \sum_{j=1}^{4} \det\left(\overline{p_j^*}, \overline{p_{j+1}^*}\right),
\quad\overline{p_{5}^*}:=\overline{p_{1}^*}. 
\end{equation}
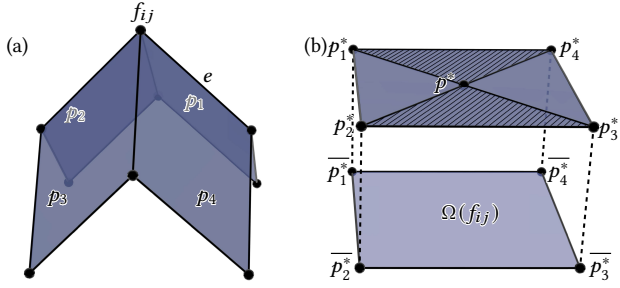
\begin{figure}[t]
\centering
\begin{overpic}
[width=0.55\linewidth]{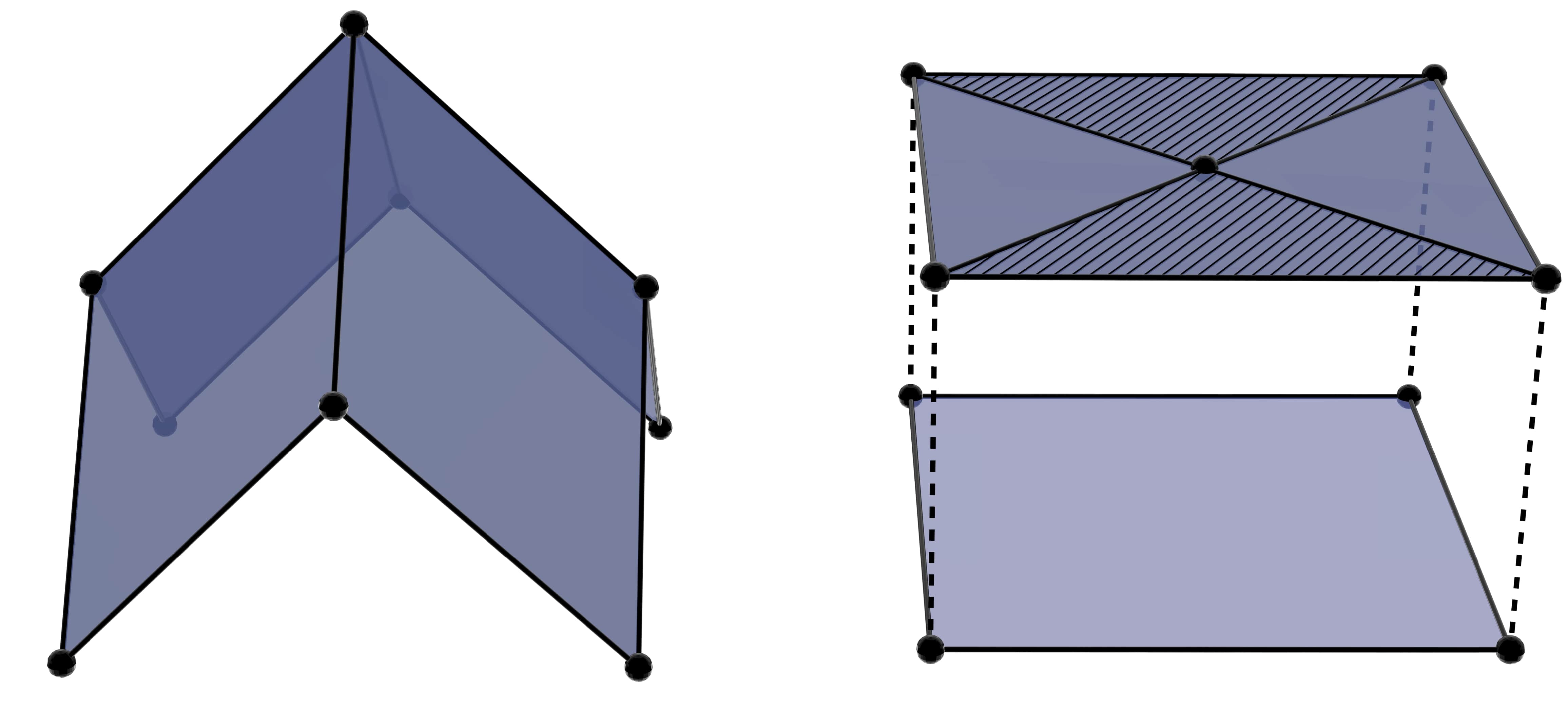}
\small
\put(0,40){\contour{white}{(a)}}
\small
\put(50,40){\contour{white}{(b)}}
\small
\put(7,15){\contour{white}{$p_3$}}
\small
\put(32,15){\contour{white}{$p_4$}}
\small
\put(10,29){\color{black!50}\contour{white}{$p_2$}}
\small
\put(30,31){\color{black!50}\contour{white}{$p_1$}}
\small
\put(72,34){\contour{white}{$p^*$}}
\small
\put(54,40){\contour{white}{$p_1^*$}}
\small
\put(93,40){\contour{white}{$p_4^*$}}
\small
\put(55,27){\contour{white}{$p_2^*$}}
\small
\put(99.5,26.5){\contour{white}{$p_3^*$}}
\small
\put(33,35){\contour{white}{$e$}}
\small
\put(54,19){\contour{white}{$\overline{p_1^*}$}}
\small
\put(91,19){\contour{white}{$\overline{p_4^*}$}}
\small
\put(54.5,3){\contour{white}{$\overline{p_2^*}$}}
\small
\put(98,3){\contour{white}{$\overline{p_3^*}$}}
\small
\put(73,12){\contour{white}{$\Omega(f_{ij})$}}
\small
\put(21,46){\contour{white}{$f_{ij}$}}
 \end{overpic}
 \captionsetup{width=0.92\linewidth}
\caption{
(a) Four consecutive faces around a vertex $f_{ij}$ of an $m\times n$ Q-net and their planes $p_1,{p_2},{p_3},{p_4}$. (b) Their metric duals form a quadrilateral ${p_1^*}{p_2^*}{p_3^*}{p_4^*}$ (a face of so-called \emph{metric dual net}).
The curvature $\Omega(f_{ij})$ at $f_{ij}$ is the oriented area of the top view $\overline{p_1^*}\,\overline{p_2^*}\,\overline{p_3^*}\,\overline{p_4^*}$.  The \emph{opposite ratio} of 
$f_{ij}$ with respect to the edge $e$ 
in (a) is the ratio of the areas of the shaded triangles $p^*p_1^*p_4^*$ and $p^*p_2^*p_3^*$ in (b). Here, the edge $e$ determines the order of faces to define the opposite ratio.  }
\label{figure:isotopic-gaussian-curvature}
\end{figure}
For instance, if 
the $i$-line through $f_{ij}$ is straight (and does not contain $f_{i\pm 1,j}$; see Fig.~\ref{fig:flexible-class-i-def}(a)), then 
$\overline{p_{1}^*}=\overline{p_{4}^*}$, $\overline{p_{2}^*}=\overline{p_{3}^*}$, and hence $\Omega(f_{ij})=0$. 


An \emph{isotropic isometric deformation} of an $m\times n$ Q-net $f_{ij}$ is a continuous family of $m\times n$ Q-nets $f_{ij}(t)$, where $t\in [0,1]$
and $f_{ij}(0)=f_{ij}$, satisfying 
\begin{description}
    \item[face condition:] corresponding faces of all $m\times n$ Q-nets $f_{ij}(t)$ are isotropically congruent; and    
    \item[vertex condition:] corresponding non-boundary vertices of all $m\times n$ 
    Q-nets $f_{ij}(t)$ have the same curvatures. 
\end{description}
An isotropic isometric deformation is \emph{trivial} if for each 
$t\in [0,1]$
there is an isotropic congruence $C_t$ such that $f_{ij}(t)=C_t(f_{ij})$ for each $0\leq i\leq m,0\leq j\leq n$. 
The 
Q-net $f_{ij}$ is an \emph{isotropic flexible net} 
if it has a nontrivial isotropic isometric deformation; see Fig.~\ref{fig:flexible-class-i-def}(a).


All isotropic flexible nets were classified in the series of papers \cite{pirahmad2024area,pirahmad2025}. Namely, under minor convexity assumptions, 
an $m\times n$ Q-net 
is an isotropic flexible net if and only if 
one of the following conditions holds: 
    \begin{enumerate}
    \item[\textup{(i)}] 
    $n$ lines $p_{k,0}\cap p_{k+2,0},\ldots,p_{k,n-1}\cap p_{k+2,n-1}$ lie in one isotropic plane for each $0\le k\le m-3$ or $m$ lines $p_{0,l}\cap p_{0,l+2},\ldots,p_{m-1,l}\cap p_{m-1,l+2}$ lie in one isotropic plane for each $0\le l\le n-3$;
    \item[\textup{(ii)}] any two non-boundary vertices joined by an edge 
    have equal opposite ratios with respect to the edge.
    \end{enumerate}
Here the \emph{opposite ratio} is defined 
in Fig.~\ref{figure:isotopic-gaussian-curvature} and condition~(i) on face planes $p_{ij}$ is illustrated in Fig.~\ref{fig:flexible-class-i-def}(b). 
The nets in resulting classes~(i) and~(ii) can be efficiently constructed; see examples in Figs.~\ref{fig:flexible-isotropic-i}, \ref{fig:flexible-isotropic-ii}, and \ref{fig:flexible-isotropic-projective}. 

This classification has sharp contrast to Euclidean geometry, where even for $3\times 3$ flexible nets, there is a huge number of classes.

An example of an isotropic flexible net is a \emph{generalized T-net}. It is defined by the condition that the parameter lines are planar, and one family of parameter lines lies in isotropic planes. In other words, $i$-lines are planar, and $j$-lines are (discrete) isotropic geodesics. 
Planar parameter lines are a favorite in architecture, and isotropic planes have 
advantages for supporting structures \cite{PP-2022}. Generalized T-nets 
belong to class (i); hence they are isotropic flexible nets. 
Conversely, most nets of class (i) can be obtained from a generalized T-nets by truncation of edges (the white edges in Fig.~\ref{fig:flexible-class-i-def}(b)).

Generalized T-nets include 
\emph{T-nets}, a class of flexible nets in Euclidean geometry introduced by Graf and Sauer \cite{sauer:1970} and now actively studied \cite{izmestiev2024isometric}, as a particular case, hence the name. Thus, T-nets are flexible in both Euclidean and isotropic geometries (after an appropriate Euclidean rotation). But the class of generalized T-nets is much larger: in particular, their faces are not necessarily trapezoids. 

\begin{figure}[t]
    \centering
\begin{overpic}
[width=0.5\linewidth]{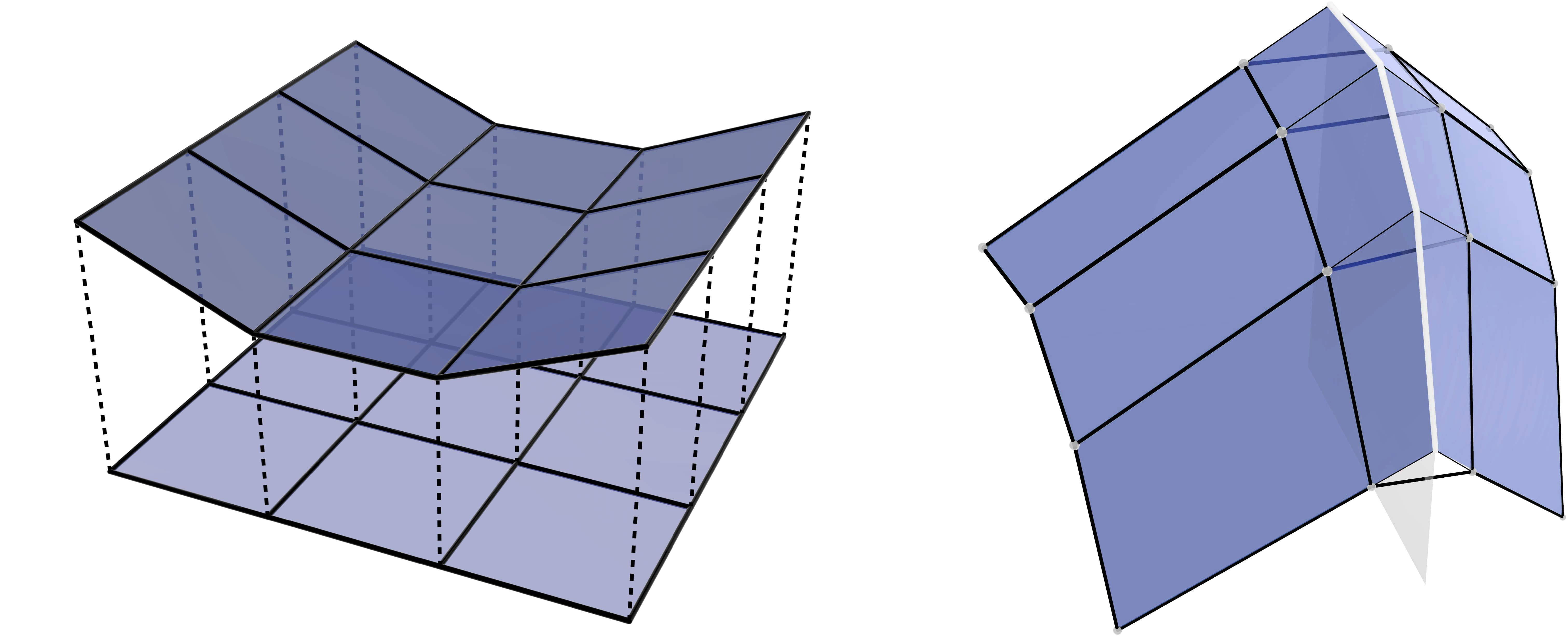}
\small
\put(0,35){\contour{white}{(a)}}
\small
\put(60,35){\contour{white}{(b)}}
 \end{overpic}
 \captionsetup{width=0.92\linewidth}
    \caption{
    (a) A planar $3\times 3$ net with square faces is flexible in isotropic geometry. Another position of the same isotropic mechanism is shown above. The top views of the vertices remain fixed (implying face condition), and $i$-lines remain straight (implying vertex condition).
    (b) An isotropic flexible 
    net of class~(i). 
    The intersections (white lines) of the planes of every other face lie in the same isotropic plane (gray). Under some convexity assumptions,
    this condition implies that the extensions (thin lines) of every other edge on $j$-lines intersect.
    Extending every other face until they meet, we construct a generalized T-net. The initial net is obtained from the resulting one by ``truncation of edges''. 
    }
    \label{fig:flexible-class-i-def}
\end{figure}

Another particular case of generalized T-nets is \emph{isotropic Voss nets}, i.e., Q-nets with all parameter lines being isotropic geodesics (see the special case in Fig.~\ref{fig:flexible-class-i-def}(a)). They are an isotropic analog of \emph{Voss nets}, a class of Euclidean flexible nets introduced by Voss \cite{sauer:1970}. 

Generalized T-nets are particularly simple to design. 

They are known to be metric dual 
to \emph{cone-cylinder nets} given by the equation
\begin{equation}
    \label{eq-p-cone-cylinder-nets}
    P_{ij}=a_i + \sigma_i b_j,
    \qquad
    0 \leq i \leq m+1\text{ and } 0 \leq j \leq n+1,
\end{equation}
for some
vectors $a_0,\dots,a_{m+1},b_0,\dots,b_{n+1}\in\mathbb{R}^3$ and real numbers $\sigma_0,\dots,\sigma_{m+1}\ne 0$. One can choose these parameters arbitrarily and apply metric duality to
get a generalized T-net \cite
{pirahmad2025}
$$f_{ij} = - \frac{1}{\det(e_3, b_{j+1}-b_j,\Delta_{ij})}
\begin{pmatrix}
    \det(e_1, b_{j+1}-b_j,\Delta_{ij}) \\
    \det(e_2, b_{j+1}-b_j,\Delta_{ij}) \\
    \det(a_{i}+\sigma_{i}b_{j}, b_{j+1}-b_j,\Delta_{ij})
\end{pmatrix},
$$
where  $e_1=(1,0,0)$, $e_2=(0,1,0)$, $ e_3=(0,0,1)$, and
$$
\Delta_{ij}:=a_{i+1}-a_i+b_j(\sigma_{i+1}-\sigma_{i}).
$$
The only restriction is a non-zero denominator
$\det(e_3, b_{j+1}-b_j,\Delta_{ij})\ne 0$. Alternatively, 
one can apply the general design method for nets with planar parameter lines from \cite[Section~3.1.2]{PP-2022}. 

\begin{figure}[tb]
\begin{overpic}
[width=1.0\linewidth]{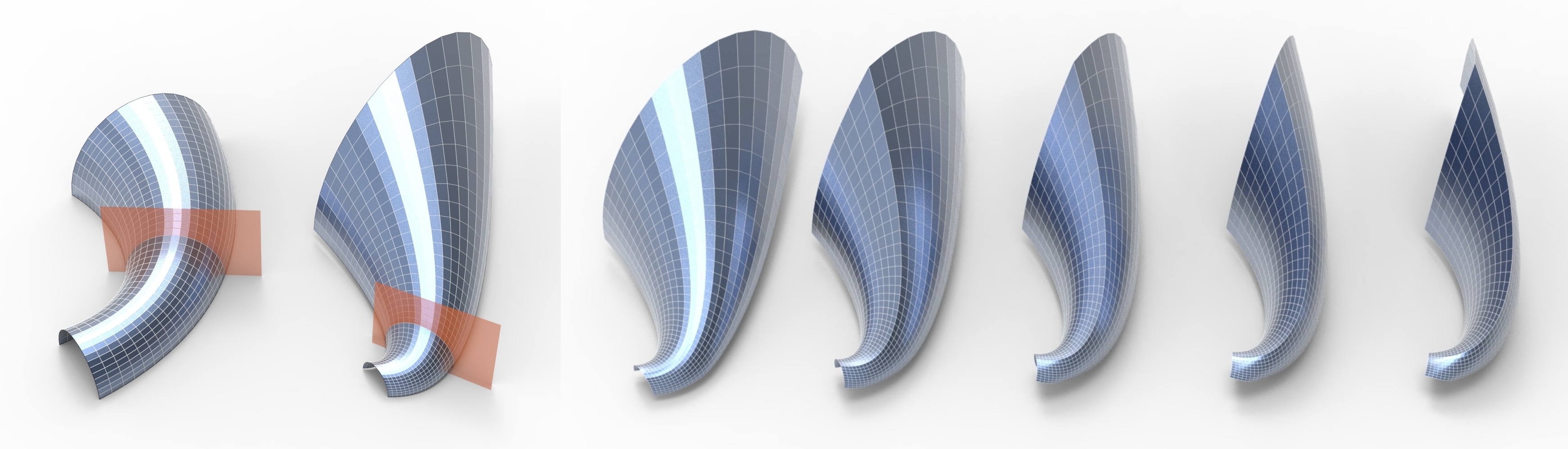}
\small
\put(1,25){\contour{white}{(a)}}
\put(16,25){\contour{white}{(b)}}
\put(36,25){\contour{white}{(c)}}
 \end{overpic}
\caption{ 
(a) A generalized T-net, flexible in isotropic geometry. The plane containing one of the parameter lines is in orange. This is a particular case of class (i) of isotropic flexible nets. Applying a projective transformation preserving the isotropic direction, we get another isotropic flexible net (b). Optimization leads to a Euclidean mechanism,
with a few positions shown in (c).
We observe a crease appearing during flexion of an originally smooth-looking mesh; cf. an analogous behavior of T-nets~\cite[Figure~18]{izmestiev2024isometric}. Such a crease may eventually lead to a collapse, but the mechanism is freely flexible within a large interval. 
}
\label{fig:flexible-isotropic-projective}
\end{figure}

With a variety of isotropic flexible nets at hand, we optimize them towards Euclidean ones using the framework of \cite{quadmech-2024}; see Fig.~\ref{fig:flexible-isotropic-i}. Surprisingly, the optimization leads to very little change in the shape. The resulting isotropic and Euclidean mechanisms have almost the same initial positions, but very different 
isometric deformations. Every position of an isotropic flexible net can be used to initialize optimization and thus leads to a different Euclidean mechanism. We can get even more Euclidean mechanisms by applying \bluer{an isotropic motion or, more generally,} a projective transformation preserving the $z$-direction (which preserves the isotropic flexibility); see Fig.~\ref{fig:flexible-isotropic-projective}. 

These results lead to several new observations. Although generalized T-nets, as well as all
isotropic flexible nets of class (i), have a family of planar parameter lines, those do, in general, not remain planar during Euclidean flexion. Therefore, we expect no new classes of Euclidean flexible $m\times n$ nets with planar parameter lines, analogous to class~(i) in isotropic geometry, which shows the advantage of our optimization approach. 
In the mechanism of Fig.~\ref{fig:flexible-isotropic-i}(b), the originally almost planar parameter lines do not remain planar during Euclidean 
flexion. 
However, for the parameter lines originally lying in isotropic planes, the deviation from planarity
during Euclidean flexion is small.
Interestingly, 
faces stay planar with high numerical precision even if we \bluer{optimize towards a mechanism with rigid, not necessarily planar faces and} do not include the planarity
of faces as objective in our optimization algorithm; \bluer{cf.~\cite{quadmech-2024}. 
Furthermore, in the mechanisms of Figs.~\ref{fig:flexible-isotropic-i}(b) and~\ref{fig:flexible-isotropic-projective}(c), we observe that all $3\times 3$ sub-nets belong to one of the \emph{equimodular} or \emph{linear elliptic/conic coupling} types of Izmestiev's classification \cite[Secs.~3.3, 3.5.4, and 3.5.5]{izmestiev-2007}. Since Izmestiev's classes form a complicated variety with components of different dimensions in the parameter space, the precision of the numerical results does not allow us to distinguish between the \emph{elliptic equimodular class} and its limiting cases, such as \emph{linear elliptic/conic coupling}. The latter seems the most probable candidate; cf.~\cite[Figure~8]{he2020rigid}. }

\newpage
\section{Asymptotic-geodesic webs} \label{sec:webs}
\begin{wrapfigure}{r}{0.23\textwidth} 
\begin{overpic}[width=1.0\linewidth]{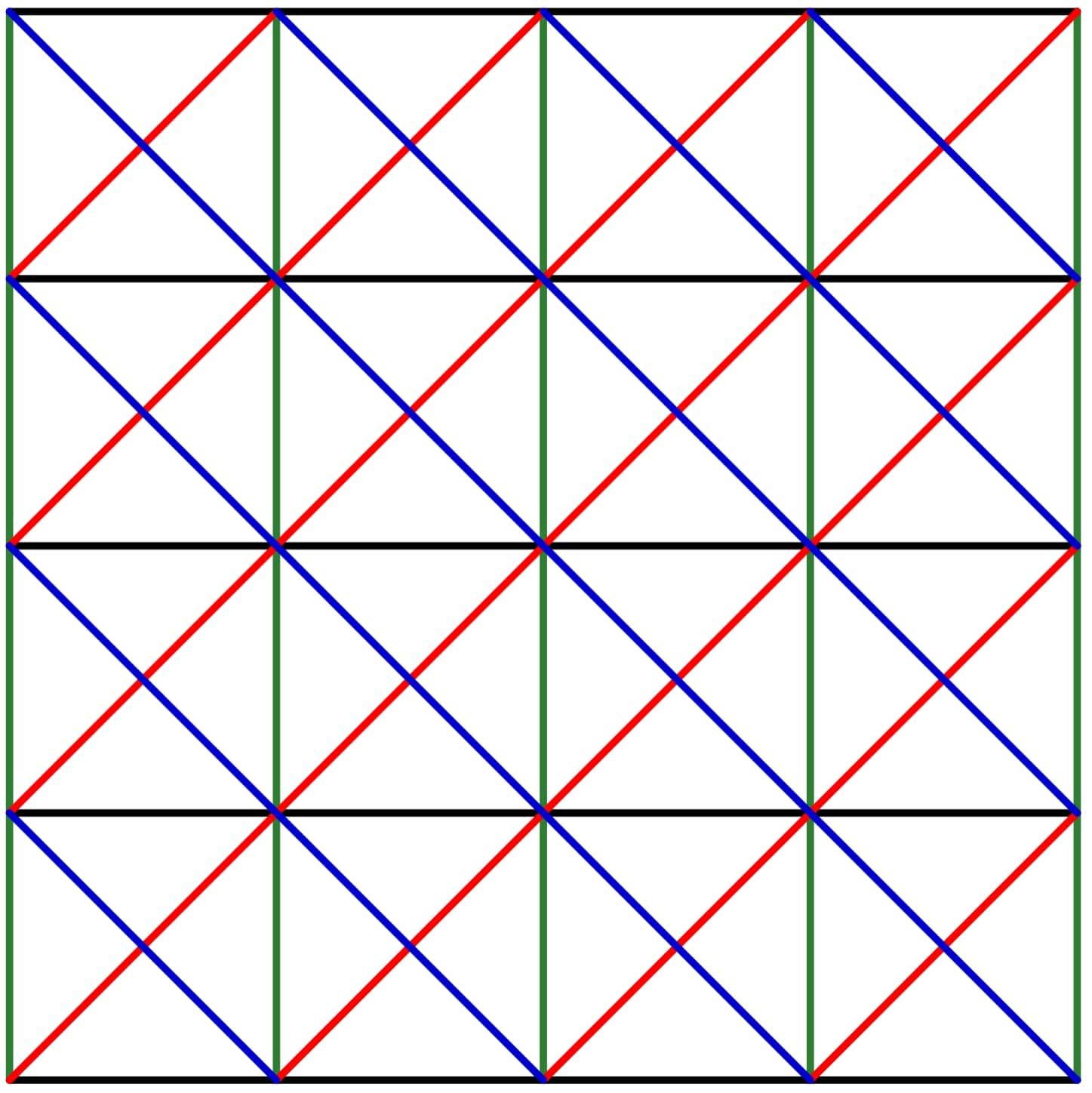}
    \small
    \put(10,-7){\contour{white}{$i = \mathrm{const}$}}
    \small
    \put(10, -12){\contour{white}{$j = \mathrm{const}$}}
    \small
    \put(56,-7){\contour{white}{$ i-j = \mathrm{const}$}}
    \small
    \put(56, -12){\contour{white}{\small $i+j = \mathrm{const}$}}
    \linethickness{1.5pt}   
    \color{darkgreen}           
    \put(0,-5){\line(1,0){7}} 
    \color{black}           
    \put(0,-10){\line(1,0){7}}  
    \color{red}           
    \put(47,-5){\line(1,0){7}}  
    \color{blue}           
    \put(47,-10){\line(1,0){7}}  
\end{overpic} 
\end{wrapfigure}

Now we present the second application of our method: the construction of asymptotic-geodesic webs.

Denote by $f\left(u,v\right)$,  a parametric representation of a smooth surface. The net of parameter lines $u = \mathrm{const}$ (\emph{u-lines}) and $v = \mathrm{const}$ (\emph{v-lines}), extended by the diagonal curves $u - v = \mathrm{const}$ (\emph{$(u - v)$-lines}), is called a \emph{smooth $3$-web (hexagonal web)} on the surface \cite{graf:sauer:1924}. Extending this $3$-web by the other diagonal curves $u + v = \mathrm{const}$ (\emph{$(u + v)$-lines}) yields a \emph{smooth $4$-web}  on the surface. 


Consider a discrete surface $f_{ij}$ of size $n \times n$. The discrete parameter lines $i = \mathrm{const}$ and $j = \mathrm{const}$  together with the diagonal polylines  joining the points $f_{ij}$ with $i - j = \mathrm{const}$ (\emph{discrete $(i - j)$-lines}), is called a \emph{discrete $3$-web}. By further extending the discrete $3$-web with the other diagonal polylines joining the points $f_{ij}$ with $i + j = \mathrm{const}$ (\emph{discrete $(i + j)$-lines}), one obtains a \emph{discrete $4$-web} (see the inset). Here all four constants as well as $i$ and $j$ are integers. The $3$- or $4$-webs are still denoted by~$f_{ij}$.

Our interest is in webs that consist of asymptotic and geodesic curves on a surface, referred to as \emph{asymptotic-geodesic webs}. We discuss three types of them, 
known as GGG, AAG, and AGAG webs \cite{web-propagation-2025, Eike-CAD-2022}. We do it first in Euclidean geometry.

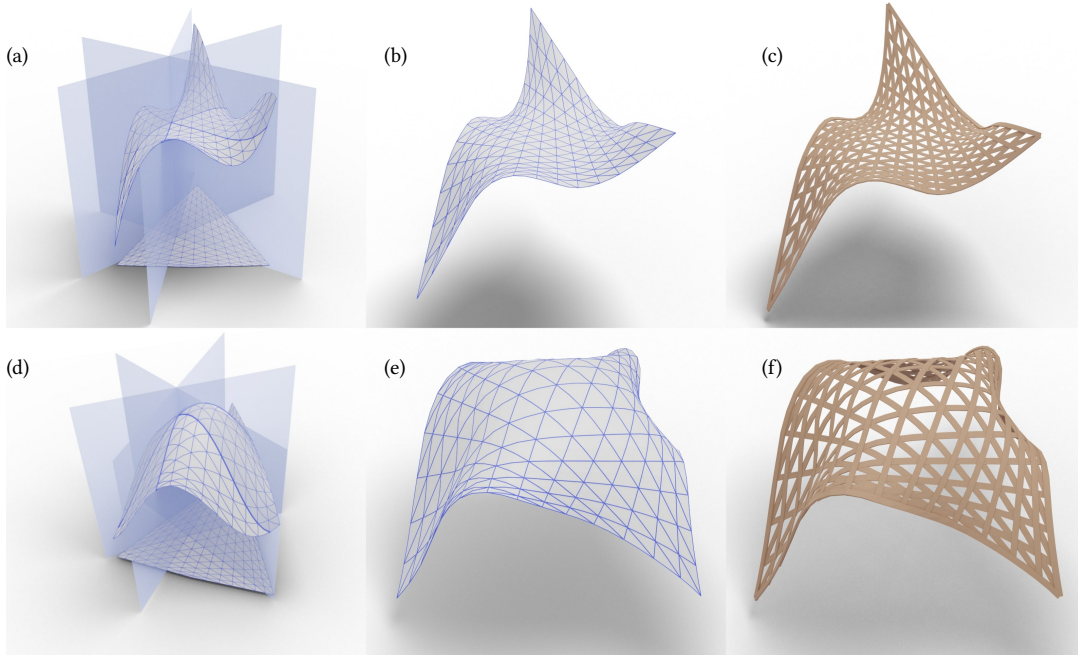
\begin{figure}[t]
    \centering
    \begin{overpic}[width=1.0\textwidth]{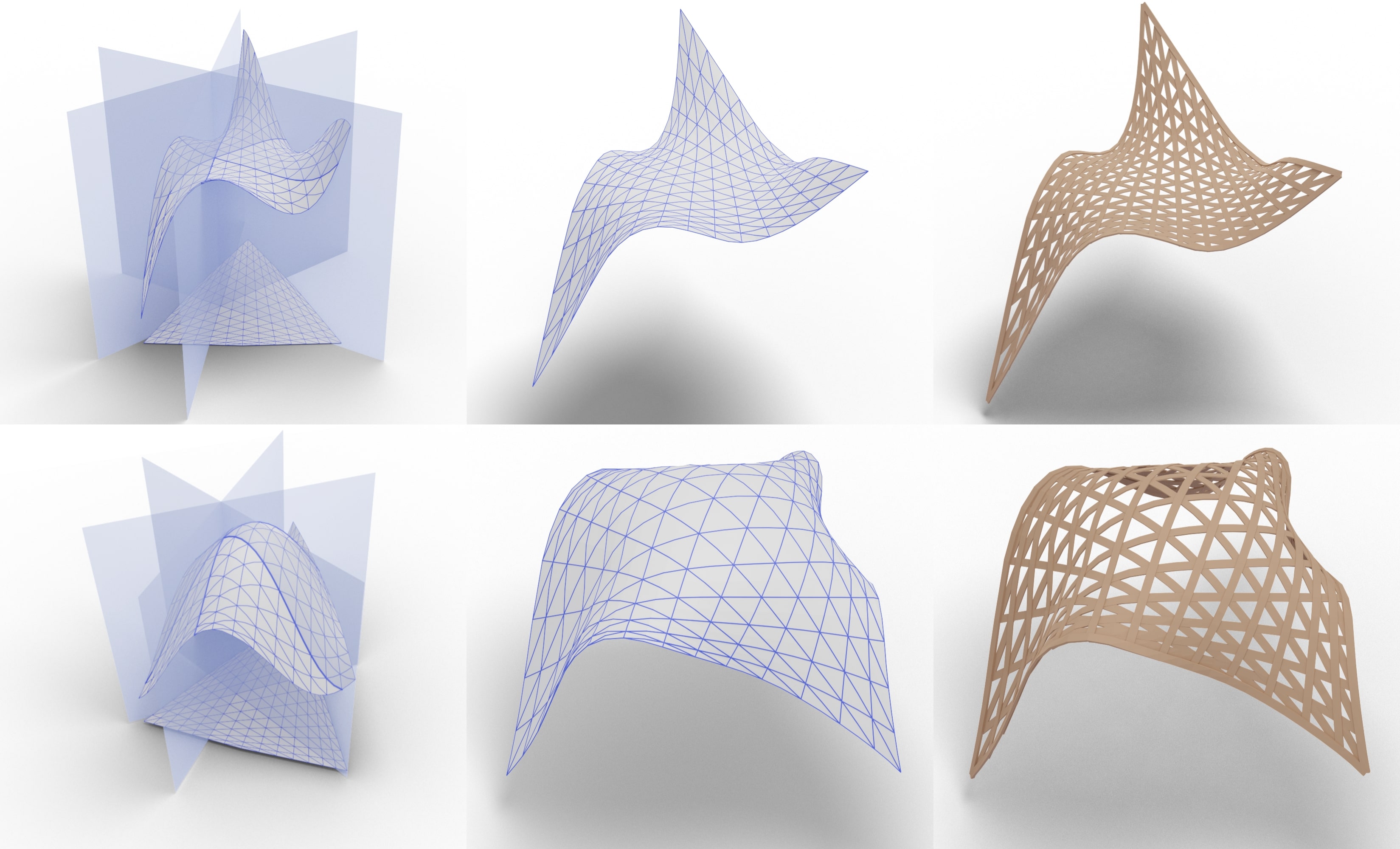} 
    \small
    \put(0,55){\contour{white}{(a)}}
    \put(35,55){\contour{white}{(b)}}
    \put(70,55){\contour{white}{(c)}}
    \put(0,26){\contour{white}{(d)}}
    \put(35,26){\contour{white}{(e)}}
    \put(70,26){\contour{white}{(f)}}
    \end{overpic}
    \caption{Design of Euclidean GGG webs by optimization of isotropic ones. 
Starting with a planar web formed by tangents to a class $3$ curve and lifting it to different surfaces, we obtain different isotropic GGG webs (a, d). The meshes were downsampled by removing some polylines to reduce
density while keeping their overall shape.  Transparent planes are isotropic and demonstrate that the blue polylines are indeed isotropic geodesics. Optimization leads to Euclidean GGG webs (b, e). We get gridshells (c, f).}
    \label{fig:GGG_merged}
\end{figure}
Recall that a \bluer{biregular curve $C(t)$ (that is, a curve such that $C'(t)$ and $C''(t)$ are linearly independent for each $t$)} on a smooth surface is  \emph{asymptotic} if the osculating plane \bluer{(spanned by $C'(t)$ and $C''(t)$)} is tangent to the surface at each point of the curve. A \bluer{biregular} curve is a \emph{geodesic} if the osculating plane is orthogonal to the surface at each point of the curve. A \emph{smooth asymptotic} (respectively, \emph{geodesic}) 
\emph{net} is a regular parametrized surface $f(u,v)$ such that the parameter lines are asymptotic curves (respectively, geodesics).

The definitions in the discrete case are aligned with these properties. Consider a discrete curve $C$ 
with the vertices $f_0, f_1, \cdots, f_k$ that are also vertices of a discrete surface $f_{ij}$. 
The edge $f_i f_{i+1}$ is called the \emph{discrete tangent} to $C$ at $f_i$ and the plane spanned by $f_{i-1}$, $f_i$, and $f_{i+1}$ is the \emph{discrete osculating plane} at $f_i$ if $i \neq 0, k$, and the three points are not collinear \cite{sauer:1970}. The unit normal $\frac{(f_i - f_{i-1}) \times (f_i - f_{i+1})}{|(f_i - f_{i-1}) \times (f_i - f_{i+1}) |}$ to this plane 
is the \emph{discrete binormal vector} at $f_i$.

A \emph{discrete normal}  vector at a non-boundary vertex $f_{ij}$  is a unit 
vector orthogonal to the two discrete tangents 
of $i$-lines and $j$-lines at $f_{ij}$, and the \emph{discrete tangent plane} 
is the one spanned by those edges.\\ Hereafter, assume that those edges indeed span a plane.



The discrete curve $C$ is a \emph{discrete asymptotic curve} 
if, at each point \bluevar{$f_i$} of~$C$ besides the endpoints and the boundary points of discrete surface, \bluer{the three neighbors $f_{i-1}$, $f_{i}$, and $f_{i+1}$ are collinear or} the discrete osculating plane is the discrete tangent 
plane. 
The curve $C$ is a \emph{discrete geodesic} if, at each non-boundary point \bluevar{$f_i$} of $C$ besides the endpoints, \bluer{the three neighbors $f_{i-1}$, $f_{i}$, and $f_{i+1}$ are collinear or} the osculating plane is orthogonal to the 
tangent plane.

If the discrete parameter lines $i=\mathrm{const}$ and $j=\mathrm{const}$ are discrete asymptotic curves, then $f_{ij}$ is a \emph{discrete asymptotic net (A-net)}. Equivalently, for all $0<i,j<n$, 
the five vertices $f_{ij}, f_{i-1, j}, f_{i+1, j}$, $f_{i, j - 1}$, and $f_{i, j + 1}$ are coplanar \cite{bobenko-2008-ddg}. In the construction of A-nets below, we also impose the same condition at the boundary, i.e., for $i,j\in\{0,n\}$, with some of those five vertices omitted (although it is not required by the definition).
%

We consider the following types of webs (discrete or smooth):
\begin{description} 
\item[GGG:] A $3$-web on a surface such that its three families of curves are geodesics (discrete or smooth). 
\item[AAG:] A $3$-web on a surface such that the parameter lines $i = \mathrm{const}$ and $j = \mathrm{const}$ are asymptotic curves, and the curves $i - j = \mathrm{const}$  are geodesics. 
\item[AGAG:] A $4$-web on a surface such that the parameter lines $i = \mathrm{const}$ and $j = \mathrm{const}$ are geodesics, and the diagonal curves $i - j = \mathrm{const}$  and $i + j = \mathrm{const}$ form one (in the smooth case) or two (in the discrete case) A-nets. 
\end{description} 
Note that in the discrete case, 
we get two A-nets (instead of one), consisting of points $f_{ij}$ with $i - j$ even and 
odd respectively. The resulting two collections of points are still called A-nets, although 
different diagonal curves consist of different numbers of points. \bluer{Beware that the discrete curves $i - j = \mathrm{const}$  and $i + j = \mathrm{const}$ are \emph{not} necessarily discrete asymptotic curves on the original discrete surface; they only become such after omitting the points $f_{ij}$ with one parity of $i - j$. The assignment of curve families to parameter lines versus diagonals is motivated by the construction of the isotropic counterparts (see Sections~4.2 and~4.3).}


We aim to transform 
webs
from isotropic to Euclidean geometry through numerical optimization. Let us define the isotropic counterparts of our webs. For that, we need to restrict ourselves to \emph{admissible} surfaces (discrete or smooth), that is, the ones having no isotropic tangent planes.
Asymptotic curves are a projective geometry concept and remain the same in both Euclidean and isotropic geometry. However, geodesics are much simpler in isotropic geometry: they are just (discrete or smooth) curves on the surface whose top views are straight line segments.
One 
defines
\emph{GGG, AAG, and AGAG webs} in isotropic geometry 
by replacing the Euclidean geodesics 
with their isotropic counterparts.


\subsection{Isotropic GGG webs}
\begin{wrapfigure}{r}{0.25\textwidth} 
\vspace{-1.5cm}
\begin{overpic}[width=1.0\linewidth]{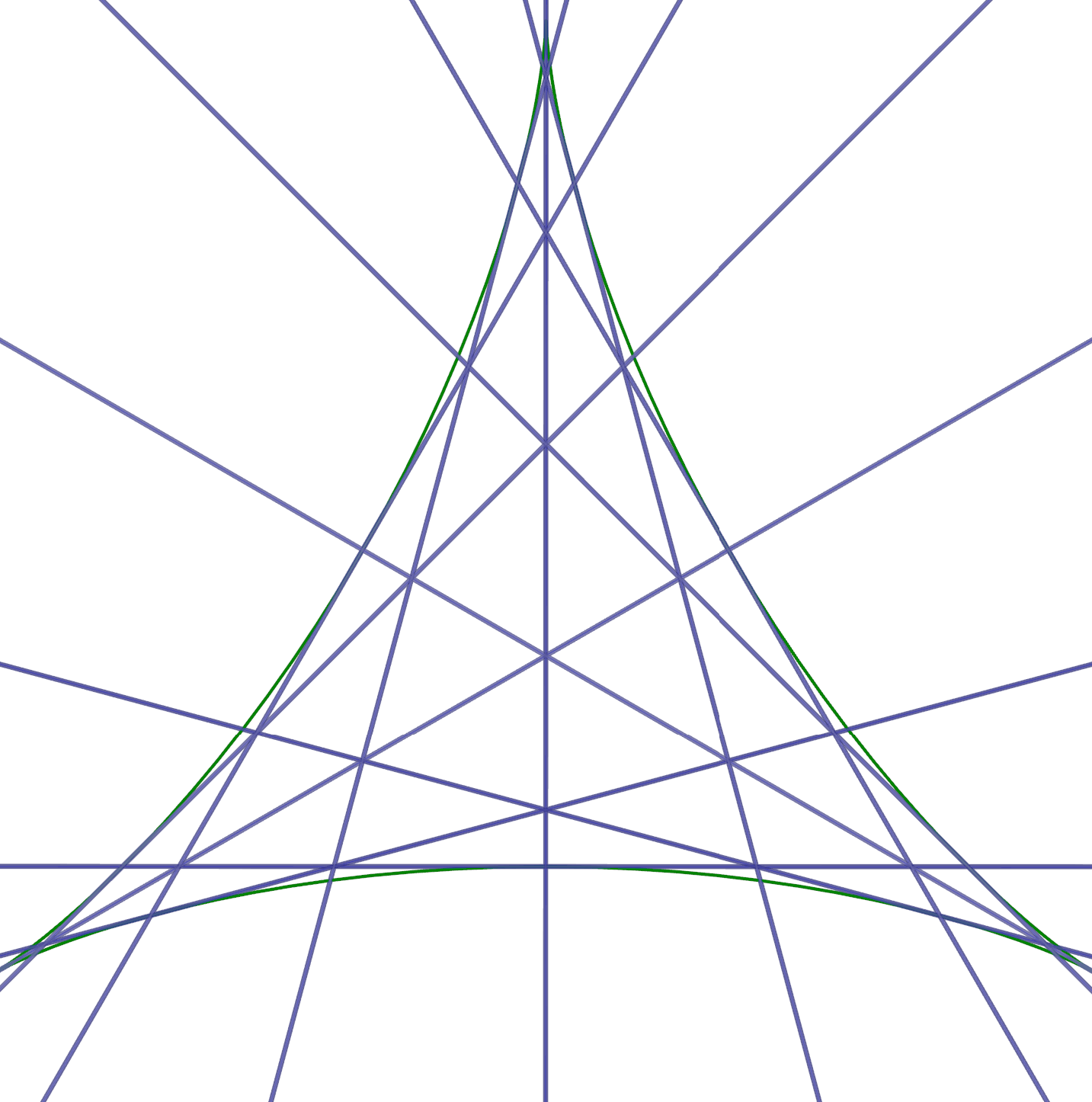}
    \small
    \put(10,-7){\contour{white}{algebraic curve of class $3$}}
    \small
    \put(10, -15){\contour{white}{$3$-web of straight lines}}
    \linethickness{1.0pt}   
    \color{darkgreen}           
    \put(0,-6){\line(1,0){7}}  
    \color{darkblue}           
    \put(0,-13){\line(1,0){7}}  
\end{overpic} 
\end{wrapfigure}
Thus, an isotropic GGG (discrete or smooth) web is a $3$-web on an admissible surface whose top view is a $3$-web of straight lines. 
Graf and Sauer \cite{graf:sauer:1924} showed that any $3$-web of straight lines is formed by the tangents of an algebraic curve of class $3$ (see the inset).
In the same work, they also provided a way to construct a discrete $3$-web of straight lines. Thus, one can obtain a discrete isotropic GGG web by projecting the  $3$-web of straight lines onto a surface, see Fig. \ref{fig:GGG_merged} (a, d).

\begin{wrapfigure}{r}{0.25\textwidth} 
\vspace{-2cm}
\begin{overpic}[width=0.8\linewidth]{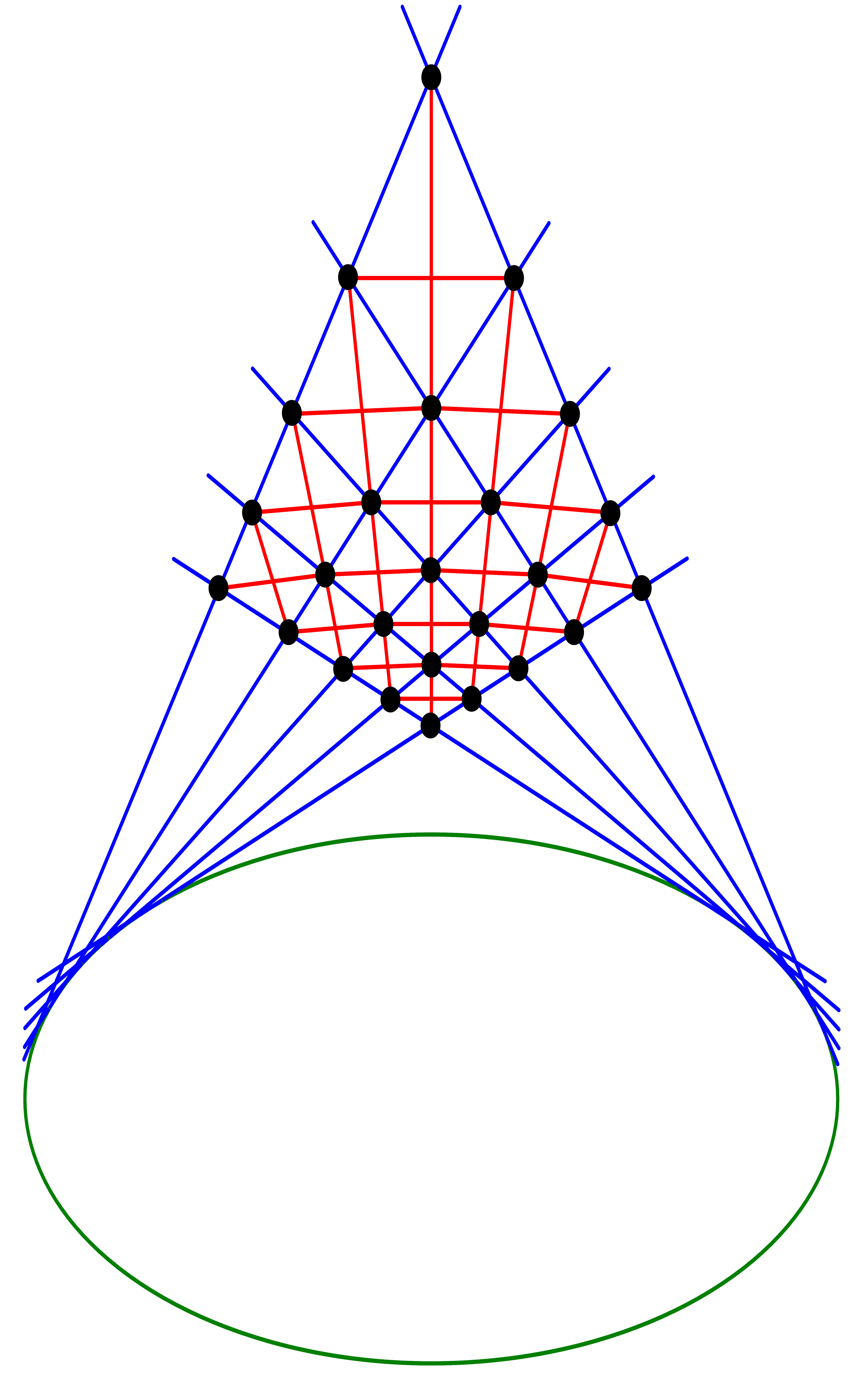}
    \small
    \put(8,-4){\contour{white}{Conic}}
    \small
    \put(8, -10.5){\contour{white}{G-net}}
    \small
    \put(38,-4){\contour{white}{A-nets}}
    \linethickness{1.0pt}   
    \color{darkgreen}           
    \put(0,-2.5){\line(1,0){5}}  
    \color{darkblue}           
    \put(0,-9){\line(1,0){5}}  
    \color{darkred}           
    \put(30,-2.5){\line(1,0){5}}  
\end{overpic} 
\end{wrapfigure}
\subsection{Isotropic AGAG webs}

\begin{figure}[t]
    \centering
    \begin{overpic}[width=1.0\textwidth]{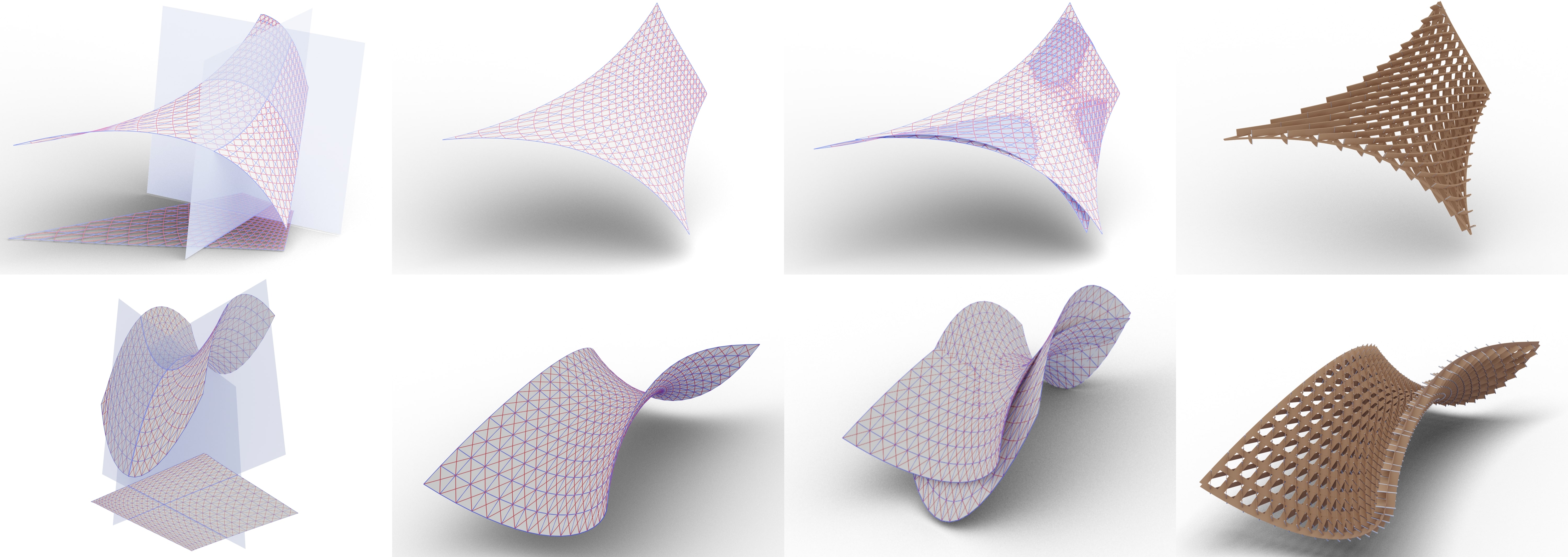} 
    \small
    \put(0,35){\contour{white}{(a)}}
    \put(26,35){\contour{white}{(b)}}
    \put(51,35){\contour{white}{(c)}}
    \put(76,35){\contour{white}{(d)}}
    \put(0,14){\contour{white}{(e)}}
    \put(26,14){\contour{white}{(f)}}
    \put(51,14){\contour{white}{(g)}}
    \put(76,14){\contour{white}{(h)}}
    \end{overpic}
    \caption{Design of Euclidean AGAG webs by optimization of isotropic AGAG webs. 
    Starting with a planar web formed by conic tangents 
    (blue) and diagonal polylines (red), we construct an isotropic AGAG web (a).  
    The blue polylines are part of the G-net, and the red polylines belong to the diagonal A-nets. The meshes were downsampled by removing some red and blue polylines 
    to reduce density. 
    Transparent planes are isotropic and demonstrate that the blue polylines are indeed isotropic geodesics.
    Optimization leads to a Euclidean AGAG web (b).
    The Euclidean (solid) and isotropic (transparent) webs are compared in (c). 
    Dropping 
    even more polylines, we extract a gridshell (d).
    Applying a projective transformation preserving the isotropic direction to (a), we get another isotropic AGAG web (e). Optimization leads to a Euclidean AGAG web (f) and a gridshell (h). } 
    \label{fig:AGAG_merged}
\end{figure}
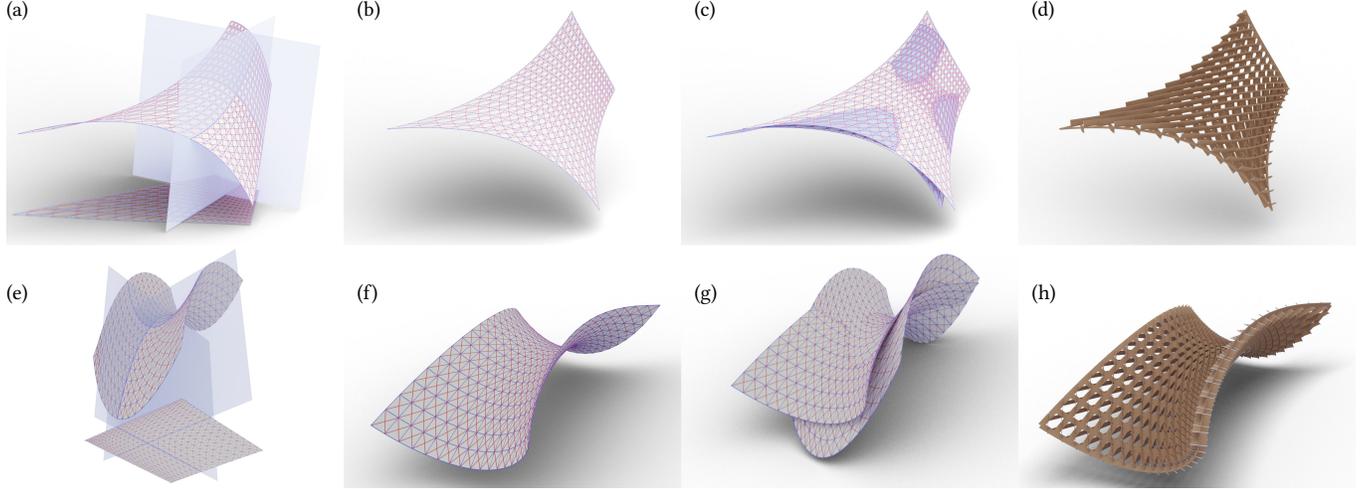

An isotropic AGAG web is a $4$-web on an admissible surface such that the $i$- and $j$-lines are geodesics (and thus form a so-called \emph{G-net}), and the two families of diagonal curves $i - j = \mathrm{const}$ and $i + j = \mathrm{const}$ form two A-nets.
Such webs were characterized by~\cite{AGAG-2024}: 
the top view of the geodesics 
lie on two families of straight lines 
tangent to the same conic (see the inset), and the A-net is constructed by lifting the two diagonal nets as described in Theorem~8 of \cite{AGAG-2024}; see Figs. \ref{fig:AGAG_merged} (a, e) and \ref{fig:AGAG1} (a). 
\subsection{Isotropic AAG Webs}

A discrete isotropic AAG web $f_{ij}$ of size $n \times n$ consists of $2n+2$ asymptotic curves ($i$- and $j$-lines) and $2n+1$ isotropic geodesics ($(i - j)$-lines) denoted by $D_0, 
\ldots, D_{2n}$ ($D_0$ and $D_{2n}$ consist of a single point each). Denote by  $\bar{D}_0, 
\ldots, \bar{D}_{2n}$ the straight lines that contain the top view of $D_0, 
\ldots, D_{2n}$, respectively. 
To design the web, we propose two methods that work for a ``general position'' input. (We omit the proofs and the definition of ``general position''; cf.~\cite{CNC-skopenkov-2020}.)

The first method is via propagation (see Algorithm~\ref{alg:AAG1}, which is run twice: for the plus and the minus sign in all ``$\pm$'', respectively). We prescribe $2n+1$ lines $\bar{D}_0, \ldots, \bar{D}_{2n}$, the points $f_{00}, f_{11}, \dots, f_{nn}$ (with the top view on $\bar{D}_n$) and $f_{0, -1}, f_{1, 0}, \dots, f_{n+1, n}$ (with the top view on~$\bar{D}_{n+1}$). Here $f_{0, -1}$ and $f_{n+1, n}$ are just auxiliary points, they are not contained in the web but are only used to determine the tangent planes at $f_{00}$ and $f_{nn}$.
For instance, one can take two space curves 
with the top views on $\bar{D}_n$ and $\bar{D}_{n+1}$, respectively, 
and pick 
$n+1$ points $f_{00}, 
\dots, f_{nn}$ and 
$n+2$ points 
$f_{0, -1}, f_{1, 0}, \dots, f_{n+1, n}$ uniformly on the curves.  

\begin{wrapfigure}{r}{0.25\textwidth} 
\vspace{-1cm}
\begin{overpic}[width=1.0\linewidth]{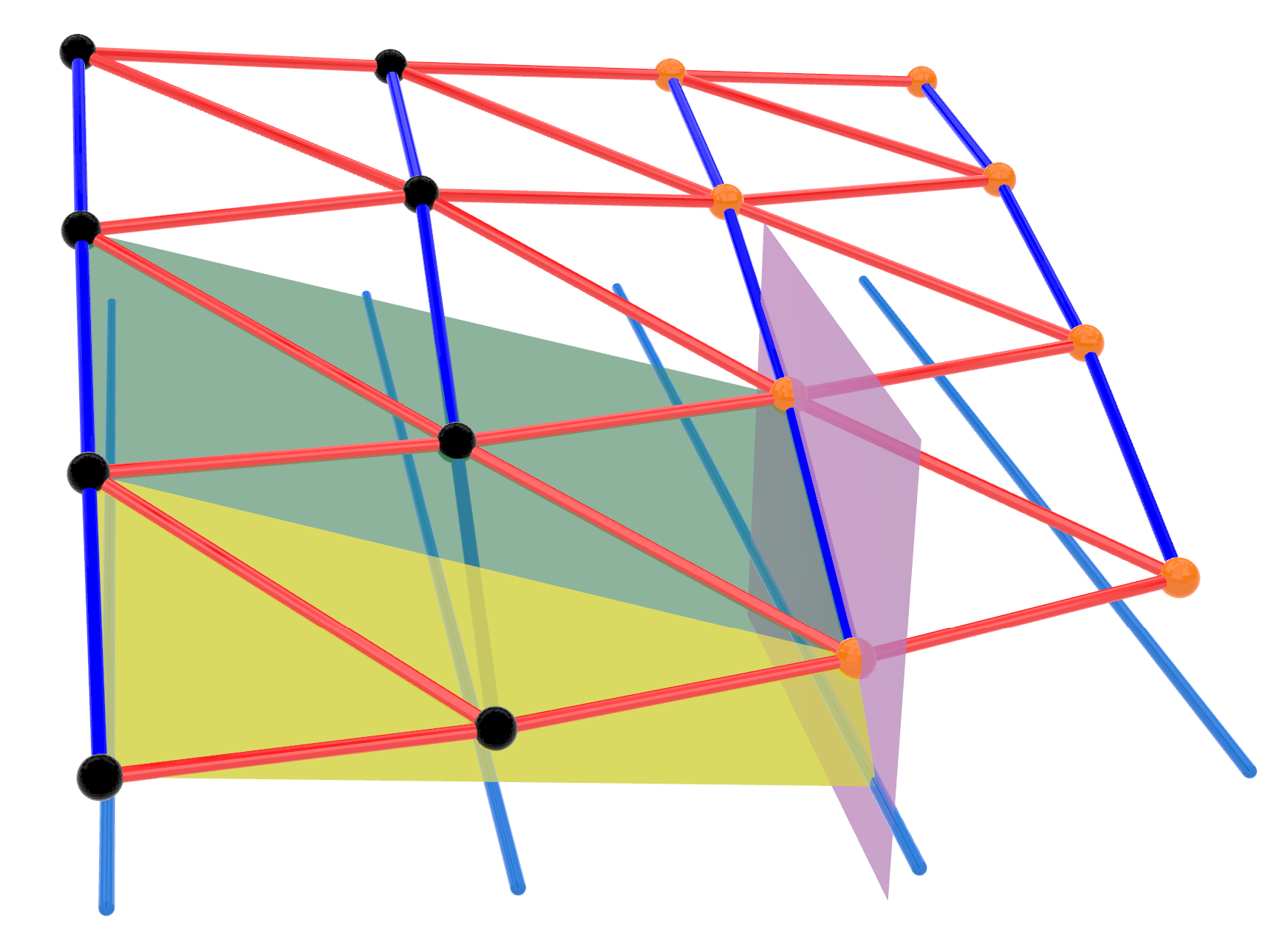}
    \small
    \put(1,7){\contour{white}{$f_{0,-1}$}}
    \small
    \put(0,32){\contour{white}{$f_{10}$}}
    \small
    \put(-1,51){\contour{white}{$f_{21}$}}
    \small
    \put(33,10){\contour{white}{$f_{00}$}}
    \small
    \put(31,32){\contour{white}{$f_{11}$}}
    \small
    \put(63,16){\contour{white}{$f_{01}$}}
    \footnotesize
    \put(9,-6){\contour{white}{Given lines}}
     \footnotesize
    \put(9,-14){\contour{white}{A-net}}
    \footnotesize
    \put(9,-22){\contour{white}{Geodesics}}
    \footnotesize
    \put(53, -6){\contour{white}{Given points}}
    \footnotesize
    \put(51, -14){\contour{white}{Constructed points}}
    \linethickness{1.0pt}   
    \color{lightblue}           
    \put(3,-4){\line(1,0){5}}  
    \linethickness{1.0pt}   
    \color{red}           
    \put(3,-12){\line(1,0){5}}  
    \color{blue}           
    \put(3,-20){\line(1,0){5}}  
    \color{black} 
    \put(50,-4){\circle*{4}}
     \color{orange} 
    \put(48,-12){\circle*{4}}
\end{overpic} 
\vspace{0.01cm}
\end{wrapfigure}
First, we construct the vertex $f_{01}$. For an A-net, it must lie on the discrete tangent planes at $f_{00}$ and $f_{11}$, and those 
planes are spanned by triples $f_{00}, f_{0, -1}, f_{1,0}$ and $f_{11}, f_{10}, f_{21}$ respectively. The top view of $f_{01}$ must lie on $\bar{D}_{n-1}$, hence $f_{01}$ lies on the isotropic plane through $\bar{D}_{n-1}$.
Thus $f_{01}$ is the intersection of these three planes; see the inset.


To construct $f_{01}$, it suffices to compute the discrete normals at $f_{00}$ and $f_{11}$; see Step~4 of Algorithm~\ref{alg:AAG1}. 

Similarly to $f_{01}$, we construct
$f_{12}, \dots, f_{n-1, n},$
using the normals along the diagonal $D_n$; see Step~$6$ of Algorithm~\ref{alg:AAG1}. 



The rest of the construction follows this logic: we determine the vertices along the diagonal $D_{n \pm i}$ using the normals along the neighboring diagonals $D_{n \pm (i-1)}$. Then, we compute the normals along $D_{n \pm i}$. We repeat this process until the entire net $f_{ij}$ is constructed. See Fig.~\ref{fig:AAG1}.
The second construction method uses planar K\oe nigs nets. 
Given  a $Q$-net  $f_{ij}$,
denote by $m_{ij}$ 
the intersection of the diagonals of the quadrilateral 
$f_{ij}f_{i,j+1}f_{i+1,j+1}f_{i+1,j}$. The net $f_{ij}$ is called a \emph{K\oe nigs} net if there exist real numbers $\nu_{ij}$, 
such that for each $0 \le i, j \le n$ 
\begin{equation}\label{nu}
\frac{f_{ij} - m_{ij}}{\nu_{ij}} = \frac{f_{i+1,j+1} - m_{ij}}{\nu_{i+1,j+1}}
\quad \text{and} \quad
\frac{f_{i,j+1} - m_{ij}}{\nu_{i,j+1}} = \frac{f_{i+1,j} - m_{ij}}{\nu_{i+1,j}}.
\end{equation}
In Theorem 8 of \cite{AGAG-2024}, it is proved that for any planar K\oe nigs net $f_{ij}$, there is an A-net whose top view 
is $f_{ij}$, and a construction of such A-net is provided.
Thus, to construct an isotropic AAG web, it suffices 
to construct a 
planar K\oe nigs net 
whose $(i-j)$-lines are straight. This is done as follows; see Algorithm~\ref{alg:AAG2}.

 \begin{wrapfigure}{r}{0.27\textwidth} 
\vspace{-0.5cm}
\begin{overpic}[width=1.0\linewidth]{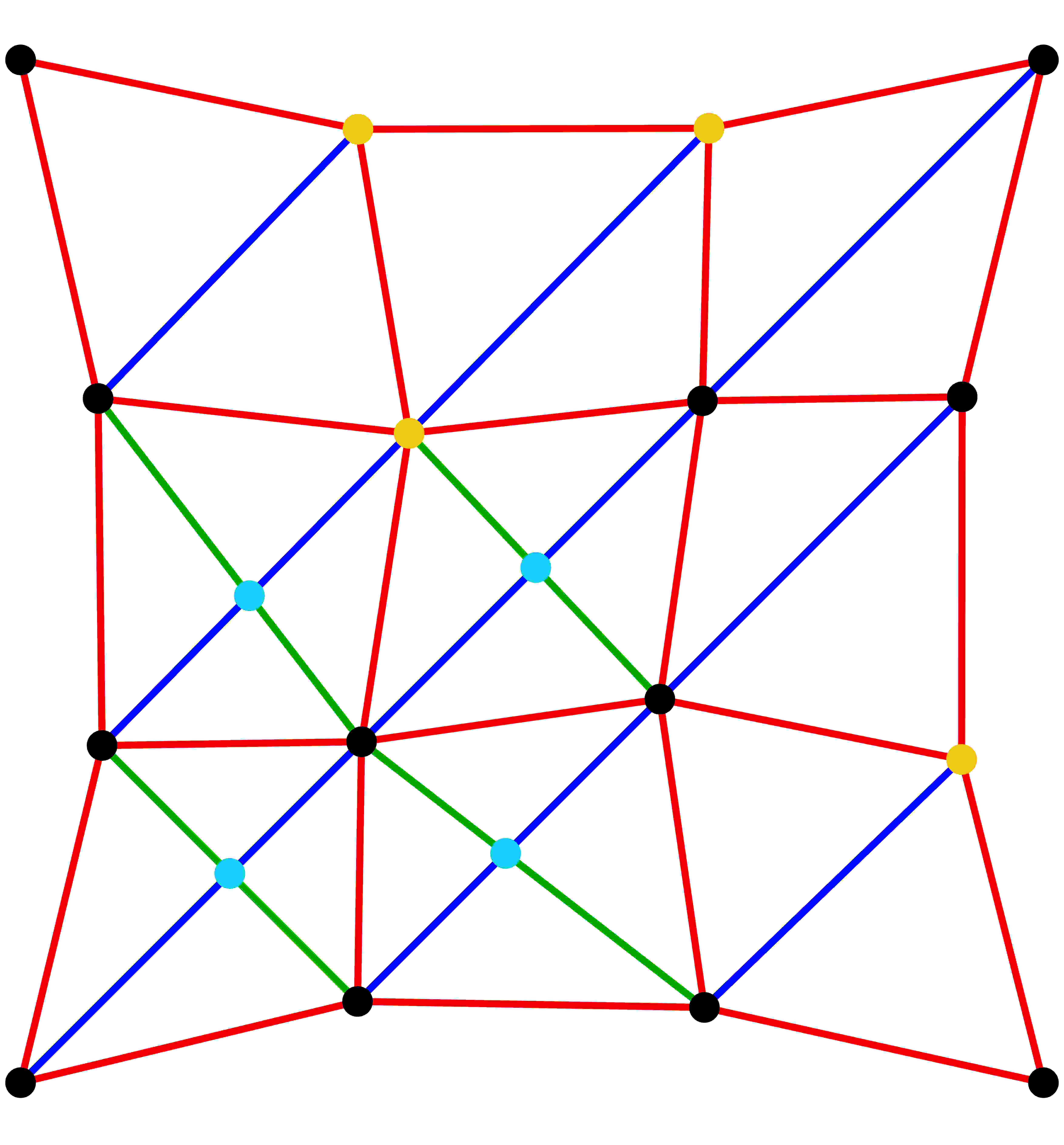}
    \small
    \put(-4,0){\contour{white}{$f_{00}$}}
    \small
    \put(30,6){\contour{white}{$f_{01}$}}
    \small
    \put(60,5){\contour{white}{$f_{02}$}}
    \small
    \put(-1,35){\contour{white}{$f_{10}$}}
    \small
    \put(-1,65){\contour{white}{$f_{20}$}}
    \small
    \put(35,65){\contour{white}{$f_{21}$}}
    \small
    \put(62,62){\contour{white}{$f_{22}$}}
    \small
    \put(60,36){\contour{white}{$f_{12}$}}
    \small
    \put(32,32){\contour{white}{$f_{11}$}}
    \small
    \put(17,17){\contour{white}{$m_{00}$}}
    \small
    \put(19,42){\contour{white}{$m_{10}$}}
    \small
    \put(41,19){\contour{white}{$m_{01}$}}
    \small
    \put(43,44){\contour{white}{$m_{11}$}}
    \scriptsize
    \put(1,-10){\contour{white}{Given lines}}
    \scriptsize
    \put(1,-17){\contour{white}{Parameter lines}}
    \scriptsize
    \put(1,-24){\contour{white}{Diagonal lines}}
    \scriptsize
    \put(48, -10){\contour{white}{Given points}}
    \scriptsize
    \put(48, -17){\contour{white}{Intersection points}}
    \scriptsize
    \put(48, -24){\contour{white}{Constructed points}}
    \linethickness{1.0pt}   
    \color{darkblue1}           
    \put(-6,-8){\line(1,0){5}}  
    \color{darkred1}           
    \put(-6,-15){\line(1,0){5}}
    \color{green}           
    \put(-6,-22){\line(1,0){5}}
    \color{black} 
    \put(46,-8){\circle*{3}}
    \color{cyan} 
    \put(46,-15){\circle*{3}}
    \color{darkyellow} 
    \put(46,-22){\circle*{3}}
\end{overpic} 
\vspace{0.1cm}
\end{wrapfigure}
We prescribe $2n+1$ diagonal lines $\bar{D}_0, \ldots, \bar{D}_{2n}$ in the plane and the boundary vertices $f_{0n}, f_{0, n-1}, \dots, f_{01}, f_{00}, f_{10}, \dots, f_{n0} $  on $\bar{D}_0, \dots, \bar{D}_{2n}$, respectively (see the inset). Additionally, we prescribe vertices $f_{11}, f_{22}, \dots, f_{nn}$ on the diagonal $\bar{D}_{n}$ and $f_{12}, f_{23}, \dots,f_{n-1, n}$ on the diagonal $\bar{D}_{n-1}$, and numbers $\nu_{00}, \nu_{01} \neq 0$. 

We construct a planar K\oe nigs net $f_{ij}$ by propagation. 
First, we find $m_{00}$ as the intersection of $\bar{D}_{n}$ with  $f_{10} f_{01}$. Then compute $\nu_{10}, \nu_{11}$ from ~\eqref{nu} for  $i = j = 0$. 

Next, we find $m_{10}$ and $m_{01}$ and compute 
 $\nu_{20}, \nu_{02}$,  and $\nu_{12}$ from ~\eqref{nu} for  $i + j = 1$.

Next, we compute the triple $(f_{21}, m_{11},\nu_{21})$ as follows: we have two equations from~\eqref{nu} for $(i,j) = (1,0)$ and $(1,1)$, and a third equation arises from the fact that $m_{11}$ 
lies on the line $\bar{D}_{n}$. We solve this system of equations at Step~$10$ of Algorithm~\ref{alg:AAG2}.
Finally, from  ~\eqref{nu} for $i = j = 1$, we get $\nu_{22}$. 
To continue, assume that the numbers $\nu_{ij}$ and the $s \times s$ net, consisting of the vertices $f_{ij}$ for $0 \leq i,j \leq s$, have been constructed. To extend it to a net of size $(s+1) \times (s+1)$, 
we find $m_{s0}$ and $m_{0s}$ 
and then compute $\nu_{s+1, 0}, \nu_{0, s+1}$ from ~\eqref{nu} for $(i, j) = (0,s)$ and $(s,0)$.  

Similarly to the triple $(f_{21}, m_{11}, \nu_{21})$ we  compute 
\[
(f_{s+1, 1}, m_{s, 1}, \nu_{s+1, 1}), \dots, (f_{s+1, s-1}, m_{s, s-1}, \nu_{s+1, s-1})
\]
and 
\[
(f_{1, s+1}, m_{1, s}, \nu_{1, s+1}), \dots, (f_{s-1, s+1}, m_{s-1, s}, \nu_{s-1, s+1})
\]
at Steps $7$ and $10$ of Algorithm~\ref{alg:AAG2}.
Then, from ~\eqref{nu} for $i = s-1$ and $j = s$, we obtain $\nu_{s, s+1}$. 
We compute
$(f_{s+1, s}, m_{s, s}, \nu_{s+1, s})$. Finally, from ~\eqref{nu} for $i = j = s$, we obtain $\nu_{s+1, s+1}$. 

Thus, the $(s+1) \times (s+1)$ net is fully constructed. Repeating this process iteratively generates the net $f_{ij}$. \bluer{We note that the stability of the propagation depends primarily on the configuration of the prescribed initial diagonal lines $\bar{D}_0, \ldots, \bar{D}_{2n}$ in the plane. To ensure regularity, it is essential to prevent intersections between the adjacent diagonal lines within the surface region. A small difference $|k_{l+1}-k_l|$ combined with a relatively large $|b_{l+1}-b_l|$ in Algorithm~\ref{alg:AAG2}  yields a sufficiently large K\oe nigs net for surface design.} See Fig.~\ref{fig:AAG21}\\

\begin{figure}[htbp]
    \centering
 \begin{overpic}
[width=0.95\linewidth]{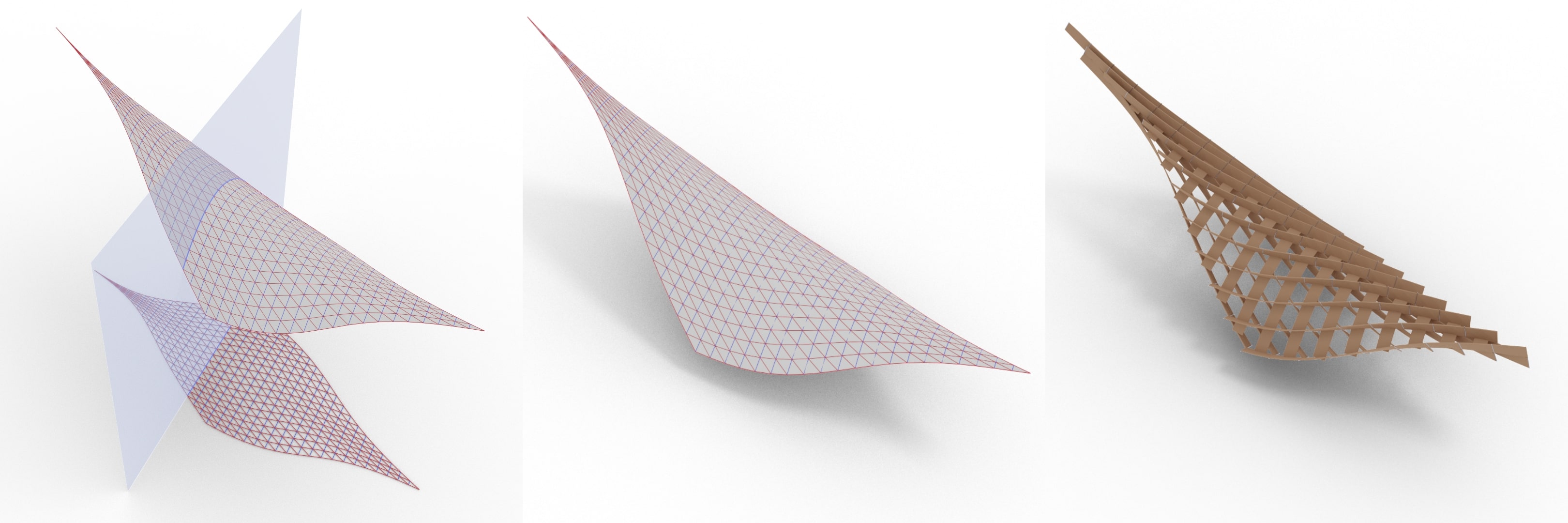}
\small
\put(0,26){\contour{white}{(a)}}
\put(33,26){\contour{white}{(b)}}
\put(67,26){\contour{white}{(c)}}
 \end{overpic}
\caption{Design of Euclidean AAG webs by optimization of isotropic ones. 
We construct an isotropic AAG web (a) by propagation (see Algorithm~\ref{alg:AAG1}).  
    Blue and red polylines represent geodesics and asymptotic curves. 
    Optimization leads to a Euclidean AAG web (b)
    and downsampling gives a gridshell (c).
    }
\label{fig:AAG1}
\vspace{-0.5cm}
\end{figure}

\begin{algorithm}[htbp]
\small{
    \caption{
    \small Design of 
    isotropic AAG  webs by propagation}
 \label{alg:AAG1}
    \KwIn{\small $2n+1$  lines $\bar{D}_l:y = k_lx + b_l$ for
    $l=0,\dots,2n$.
    Points $f_{00}, f_{11}, \cdots, f_{nn}$ (with the top view  on $\bar{D}_n$) and $f_{0, -1}, f_{1, 0}, 
    \cdots, f_{n+1, n}$ (with the top view on~$\bar{D}_{n+1}$).}
    \KwOut{A discrete isotropic AAG web $f_{ij}$ of size $n\times n$ containing the given points 
    such that the top views of $(i-j)$-lines are on the given lines 
    $\bar{D}_0, \ldots, \bar{D}_{2n}$.}
     \For{$l \gets 0$ \textbf{to} $n$}{
     \For{$i \gets 0$ \textbf{to} $n$}{
    \If{$i \pm l \le n$ and $i \pm l \ge 0$ }{
    \If{$l(l\pm 1) > 0$}{
    Compute    $f_{i, i \pm l}$ solving the system  \qquad
    $\begin{cases}
    \mathbf{n}_{i, i \pm (l-1)} \cdot (f_{i, i \pm l} - f_{i, i \pm (l-1)}) = 0,\\
    \mathbf{n}_{i \pm 1, i \pm l} \cdot (f_{i, i \pm l} - f_{i\pm 1, i \pm l}) = 0,\\
    y_{i, i \pm l}  = k_{n \mp l} x_{i, i \pm l} +b_{n \mp l} \, ;
    \end{cases}$}
    \If{$l\pm 1 \ge 0$}{
    $\mathbf{n}_{i, i \pm l} \gets (f_{i, i \pm l} - f_{i, i \pm (l -1)}) \times (f_{i, i \pm l} - f_{i \pm 1, i \pm l}) $ \;
    }}}
    }
    Return the web $f_{ij}$ \;}
    \end{algorithm}

\begin{figure}[htbp]
\vspace{-1.5cm}
    \centering
 \begin{overpic}
[width=0.55\linewidth]{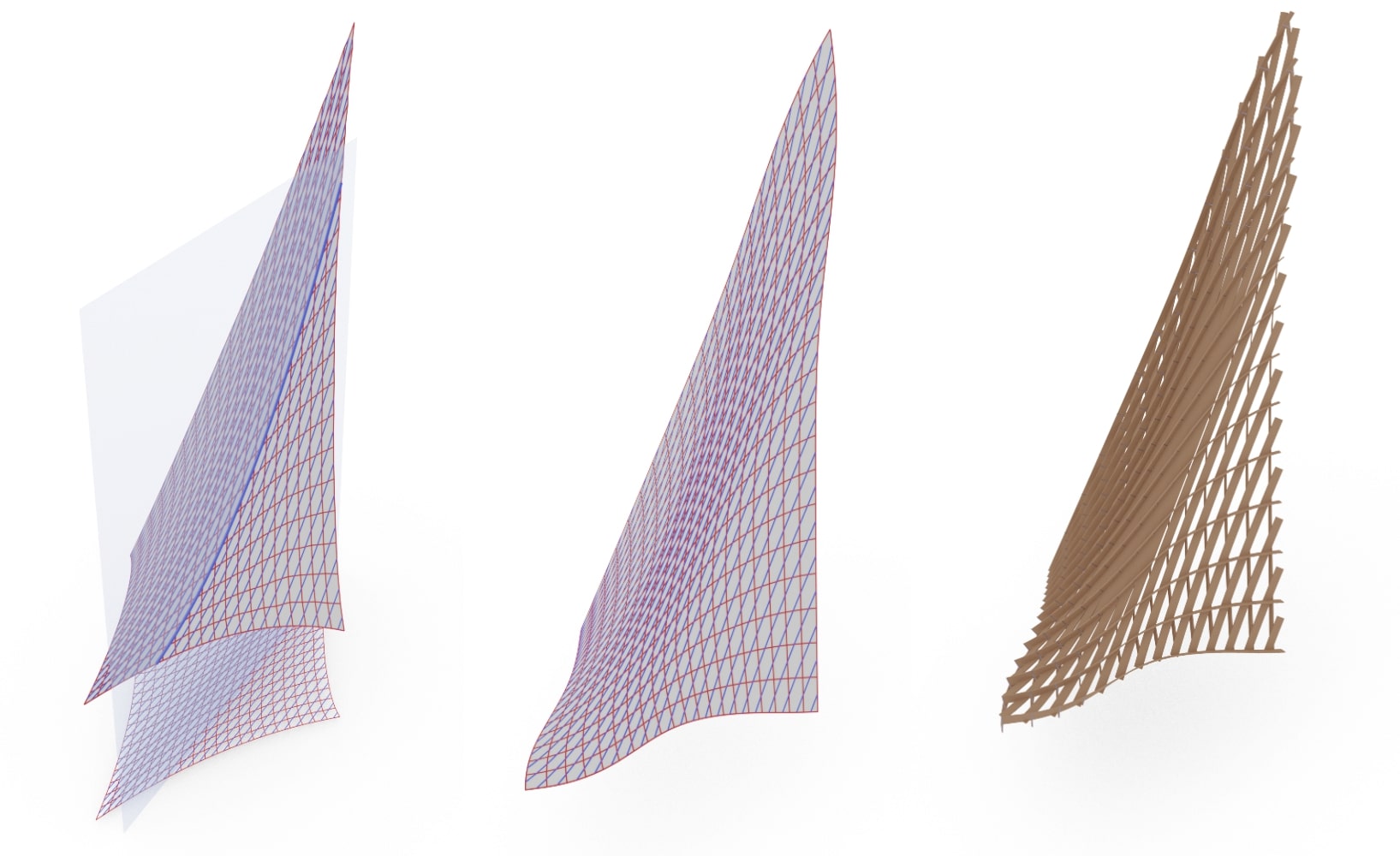}
\small
\put(6,55){\contour{white}{(a)}}
\put(41,55){\contour{white}{(b)}}
\put(78,55){\contour{white}{(c)}}
 \end{overpic}
\caption{\small{Another design method for Euclidean AAG webs based on Algotithm~\ref{alg:AAG2}.
Starting with a planar 
K\oe nigs net (red) with one family of diagonals forming straight lines (blue), we construct an isotropic AAG web (a). Optimization leads to a Euclidean AAG web (b) and a gridshell (c). 
}}
\label{fig:AAG21}
%
\end{figure}
\begin{algorithm}[htbp]
    \caption{
    \small Design of 
    planar K\oe nigs nets}
 \label{alg:AAG2}
    \small \KwIn{  $2n+1$ lines $\bar{D}_l$ 
    : $y = k_lx + b_l$ for $l=0,\dots,2n$. Points $f_{0, n}, f_{0, n-1}, \cdots, f_{0, 0}$,  $f_{1,0}, f_{2, 0}, \cdots, f_{n, 0}$ on $\bar{D}_0, \ldots, \bar{D}_{2n}$, respectively, $f_{00}, f_{11}, \cdots, f_{nn}$ on $\bar{D}_{n}$ , and $f_{0, 1}, f_{1, 2}, \cdots,f_{n-1, n}$ on $\bar{D}_{n-1}$ and numbers $\nu_{00}, \nu_{01} \ne 0$. }
    \small \KwOut{ A planar K\oe nigs net $f_{ij}$ of size $n\times n$ containing the given points 
    and such that $(i-j)$-lines are on the given lines 
    $\bar{D}_0, \ldots, \bar{D}_{2n}$}
    \small{
    \For{$s \gets 0$ \textbf{to} $n-1$}{
      $m_{s, 0} \gets \bar{D}_{n + s} \cap f_{s+1, 0}f_{s, 1}$ and $m_{0, s} \gets \bar{D}_{n - s} \cap f_{0, s+1}f_{1, s}$ \;
      $\nu_{s+1, 0} \gets \nu_{s,1}(f_{s+1, 0} - m_{s,0}) / (f_{s,1} - m_{s,0})$ \;
      $\nu_{0, s+1} \gets \nu_{1,s}(f_{0, s+1} - m_{0,s}) / (f_{1,s} - m_{0,s})$ \;
    \For{$i \gets 1$ \textbf{to} $s$}{
    \If{$i < s$ }{
    Compute  $f_{i, s+1}$,
    $m_{i, s} = (x_{i,s}, y_{i,s})$,  $\nu_{i, s+1}$   from the system 
    $\begin{cases}
       \nu_{i-1, s}(f_{i, s+1} - m_{ i-1, s})=\nu_{i, s+1}(f_{i-1,s} - m_{i-1,s}),\\
        \nu_{i+1, s}(f_{i, s+1} - m_{i, s})  = \nu_{i, s+1}(f_{i+1, s} - m_{i, s}) ,  \\
        y_{i, s}  = k_{n + i- s}x_{i, s} + b_{n + i- s} \, ;
    \end{cases}$}
     \Else{$\nu_{i, s+1} \gets \nu_{i-1,s}(f_{i, s+1} - m_{i-1,s}) / (f_{i-1,s} - m_{i-1,s})$ \;}
    Compute  $f_{s+1, i}$,
    $m_{s, i} = (x_{s,i}, y_{s,i})$,  $\nu_{s+1, i}$   from the system 
    $\begin{cases}
       \nu_{s, i-1}(f_{s+1, i} - m_{s, i-1})=\nu_{s+1, i}(f_{s, i-1} - m_{s, i-1}),\\
        \nu_{s, i+1}(f_{s+1, i} - m_{s, i})  = \nu_{s+1, i}(f_{s, i+1} - m_{s, i}) ,  \\
        y_{s, i}  = k_{n + s- i}x_{s, i} + b_{n + s- i} \, ;
    \end{cases}$}  
    $\nu_{s+1, s+1} \gets \nu_{s,s}(f_{s+1, s+1} - m_{s,s}) / (f_{s,s} - m_{s,s})$ \;}
    }
    Return the web $f_{ij}$\;
     \end{algorithm}

\newpage
\section{Asymptotic nets with a constant angle} \label{sec:CRPC}



Now we turn to the third application of our 
approach: construction of 
surfaces with a constant angle 
between asymptotic curves. 



We already know what to do whenever we face a hard Euclidean problem: consider its isotropic analog, that is, the surfaces with a constant \emph{isotropic} angle $\gamma$ between asymptotic curves. 
This is equivalent to the constant ratio $H^2/K=-\cot^2\gamma$.
See \cite{Yorov-etal}. 

It is striking that there is a simple approximate analytic expression for such \emph{isotropic \bluer{CRPC} surfaces} when the angle $\gamma$ is not too far from $90^\circ$ (nothing like that is available in the Euclidean case). 

The formula is especially elegant if one uses the complex coordinate $w=x+iy$ in the $xy$-plane. Using the complex derivatives $f_w:=\frac{1}{2}\left(f_x-if_y\right)$ and $f_{\bar w}:=\frac{1}{2}\left(f_x+if_y\right)$, we 
write the isotropic mean and Gaussian curvatures of a surface $z=f(w)$
as $H=2f_{w\bar w}$ and $K=4f_{w\bar w}^2-4f_{ww}f_{\bar w\bar w}$. The equation $H^2/K=-\cot^2\gamma=\mathrm{const}$ takes the form (up to an overall sign)
\begin{equation}\label{eq-complex-crpc}
f_{w\bar w}=\cos\gamma \cdot\sqrt{f_{ww}f_{\bar w\bar w}}.
\end{equation}
\begin{figure}[b]
    \centering
 \begin{overpic}
[width=1.0\linewidth]{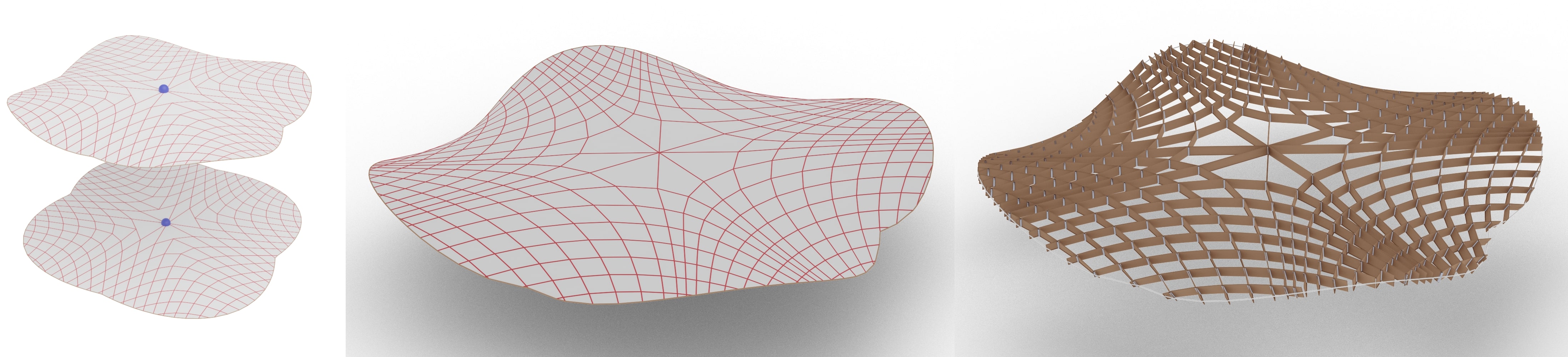}
\small
\put(0,20){\contour{white}{(a)}}
\put(25,20){\contour{white}{(b)}}
\put(64,20){\contour{white}{(c)}}
 \end{overpic}
\caption{Design of a Euclidean CRPC surface with a prescribed flat point (blue dot) and a $60^\circ$ angle between asymptotic curves,
by optimizing a second-order approximation of the isotropic CRPC surface. The initial shape is constructed using Algorithm~\ref{alg:C2L}. 
It is then remeshed using the libigl~\cite{libigl} implementation of mixed-integer quadrangulation~\cite{quadrangulation}. The remeshed surface, including its top view, is shown in (a). Optimization leads to a Euclidean CRPC surface (b) and a gridshell (c).\\}
\label{fig:CRPC3}
\end{figure}
\begin{algorithm}[bt]
    \caption{
    Design of an isotropic \bluevar{CRPC} surface with given flat points (where more than two asymptotic curves meet)} 
 \label{alg:C2L}
    \small \KwIn{
    the isotropic angle $\gamma$ between the asymptotic curves,
    the top views $w_1,\dots,w_n$ of the flat points}
   \small \KwOut{A function $f^{\mathrm{appr}}(w)$ 
    whose graph has approximately constant isotropic angle $\gamma$
    between the asymptotic curves 
    and flat points near $w_1,\dots,w_n$}
    \small{Compute $h(w):=\int (w-w_1)\dots (w-w_n)\,dw$ (expand and integrate term-wise)
    \;
    Expand $(w-w_1)^2\dots (w-w_n)^2$ and integrate term-wise twice to get a polynomial $g(w)$ such that $g''(w)=h'(w)^2$\;
    Set $f^{(0)}(w):=2\mathrm{Re}\,g(w)$, \quad $f^{(1)}(z):=|h(w)|^2$, \quad
    $\varepsilon:=\cos\gamma$, \quad
    $f^{(2)}(z):=2\mathrm{Re}\, \left(h(w)^2\right) \log(|h'(w)|+\varepsilon)$\;
    Return $f^{\mathrm{appr}}(w):=f^{(0)}(w)+f^{(1)}(w)\varepsilon+f^{(2)}(w)\varepsilon^2/2$.}
 \end{algorithm}

\begin{figure}[b]
    \centering
 \begin{overpic}
[width=1.0\linewidth]{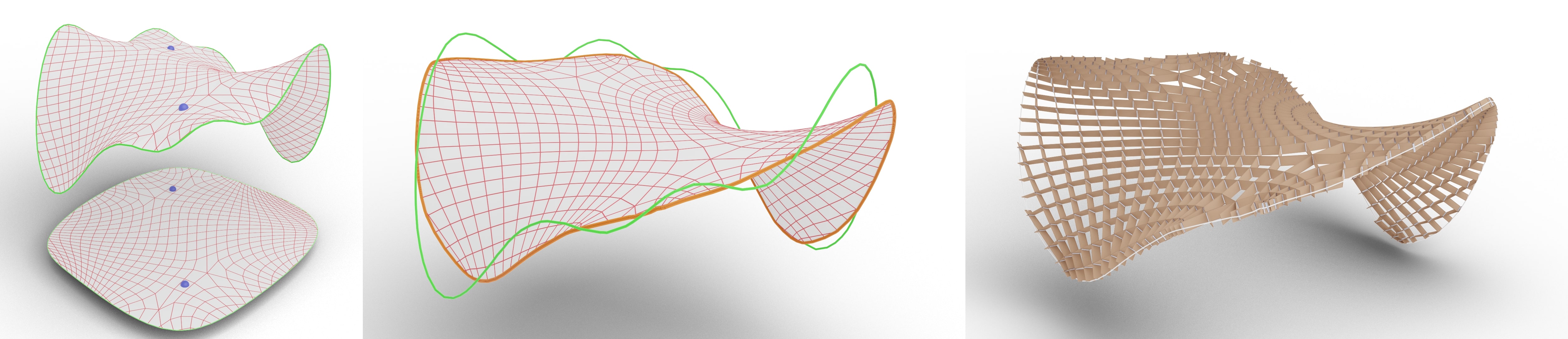}
\small
\put(-1,20){\contour{white}{(a)}}
\put(24,20){\contour{white}{(b)}}
\put(60,20){\contour{white}{(c)}}
 \end{overpic}
\caption{ Design of an Euclidean CRPC surface with a given boundary (green), two flat points (blue), and an angle of $70^\circ$ between asymptotic curves.
The initial shape, constructed using Algorithm~\ref{alg:boundary+features} and remeshed, is shown with its top view in (a). Each of the 
flat points (blue) has split into a pair of singular vertices due to limitations of the remeshing algorithm. 
Optimization leads to a Euclidean CRPC surface (b) and a gridshell~(c).}
\label{fig:CRPC2}
\end{figure}

\begin{algorithm}[t]
    \caption{
   Design of an approximate isotropic \bluevar{CRPC} surface with a given boundary 
   and flat points. 
   }
   \label{alg:boundary+features}
   \small{\KwIn{
    The isotropic angle $\gamma$ between the asymptotic curves,
    the top views $w_1,\dots,w_n$ of some of the flat points, 
    a 
    function $b(w)$ on the boundary 
    of a Jordan domain~$\Omega$, 
    the approximation order $k$
    }
    \KwOut{A function $f^{\mathrm{appr}}(w)$ in $\Omega$,
    whose graph has 
    nearly constant isotropic angle $\gamma$
    between the asymptotic curves, 
    flat points near $w_1,\dots,w_n$ (and maybe other),
    and boundary close to the graph of $b(w)$. 
    }
    Introduce the function $f^{\mathrm{appr}}(h_0,\dots,h_k,g_0,g_1;w):=2\mathrm{Re}\,g(w)+\varepsilon\cdot|h(w)|^2+\varepsilon^2\cdot \mathrm{Re}\, \left(h(w)^2\right) \log\left(|h'(w)|+\varepsilon\right)$, where 
    $h(w):=\int (w-w_1)\cdots(w-w_n)(h_0+h_1w+\cdots+h_kw^k)\,dw$, 
    $g(w):=\int\left(\int h'(w)^2\,dw\right)\,dw+g_0+g_1w$,  and $\varepsilon:=\cos\gamma$
    \; 
    Find $h_0,\dots,h_k,g_0,g_1$ that minimize $\|f^{\mathrm{appr}}(h_0,\dots,h_k,g_0,g_1;w)-b(w)\|$ on $\partial\Omega$ by optimization\;
    Return the resulting function $f^{\mathrm{appr}}(h_0,\dots,h_k,g_0,g_1;w)$}
\end{algorithm}

Let us search for a solution in the form of 
series in the small parameter $\varepsilon=\cos\gamma$:
\begin{equation}\label{eq-taylor}
f(w)=f^{(0)}(w)+ f^{(1)}(w)\varepsilon+ f^{(2)}(w)\varepsilon^2/2+\dots,
\end{equation}
for some unknown functions $f^{(0)}(w),f^{(1)}(w),
\dots$. 
Here $f^{(0)}(w)$ satisfies~\eqref{eq-complex-crpc} for 
$\varepsilon=\cos\gamma=0$, hence is a harmonic function.
Any harmonic function (in a simply-connected domain) can be written as
\begin{equation}\label{eq-f0}
    f^{(0)}(w)=2\mathrm{Re}\,g(w)
\end{equation}
for some complex analytic function $g(w)$ (roughly, a polynomial or 
power series in 
$w$). For an 
analytic function, 
$g_{\bar w}(w)=0$ and we denote $g_{w}(w)=:g'(w)$. The next function $f^{(1)}(w)$ satisfies the equation
$$
f^{(1)}_{w\bar w}=\sqrt{f^{(0)}_{ww}f^{(0)}_{\bar w\bar w}}.
$$
The right side equals $|g''(w)|$,
and it is not hard to guess a particular solution
\begin{equation}\label{eq-f1}
f^{(1)}(w)=|h(w)|^2, \quad\text{ where }h(w):=\int\sqrt{g''(w)}\,dw.
\end{equation}
Technically, we need some assumptions to get a continuous branch of the square root here; 
e.g., 
the requirement that all zeroes of $g''(w)$, if any, have multiplicity $2$.  
We present the details (unessential for implementation) in a subsequent publication~\cite[Section~5]{SkopenkovYorov2025CRPC}.


\begin{figure}[t]
    \centering
 \begin{overpic}
[width=0.4\linewidth]{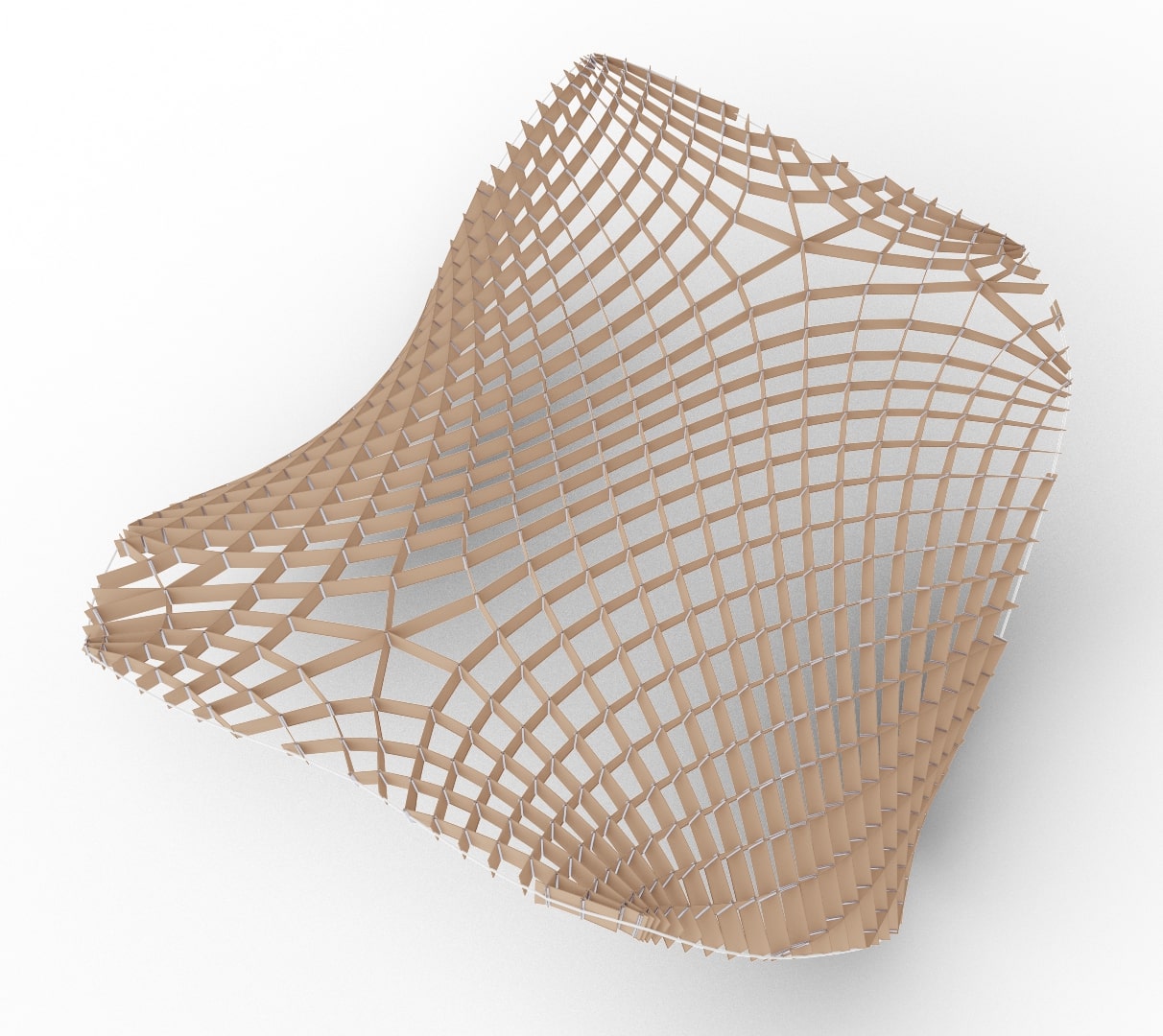}
\small
\put(0,80){\contour{white}{(a)}}
 \end{overpic}
 \begin{overpic}
[width=0.55\linewidth]{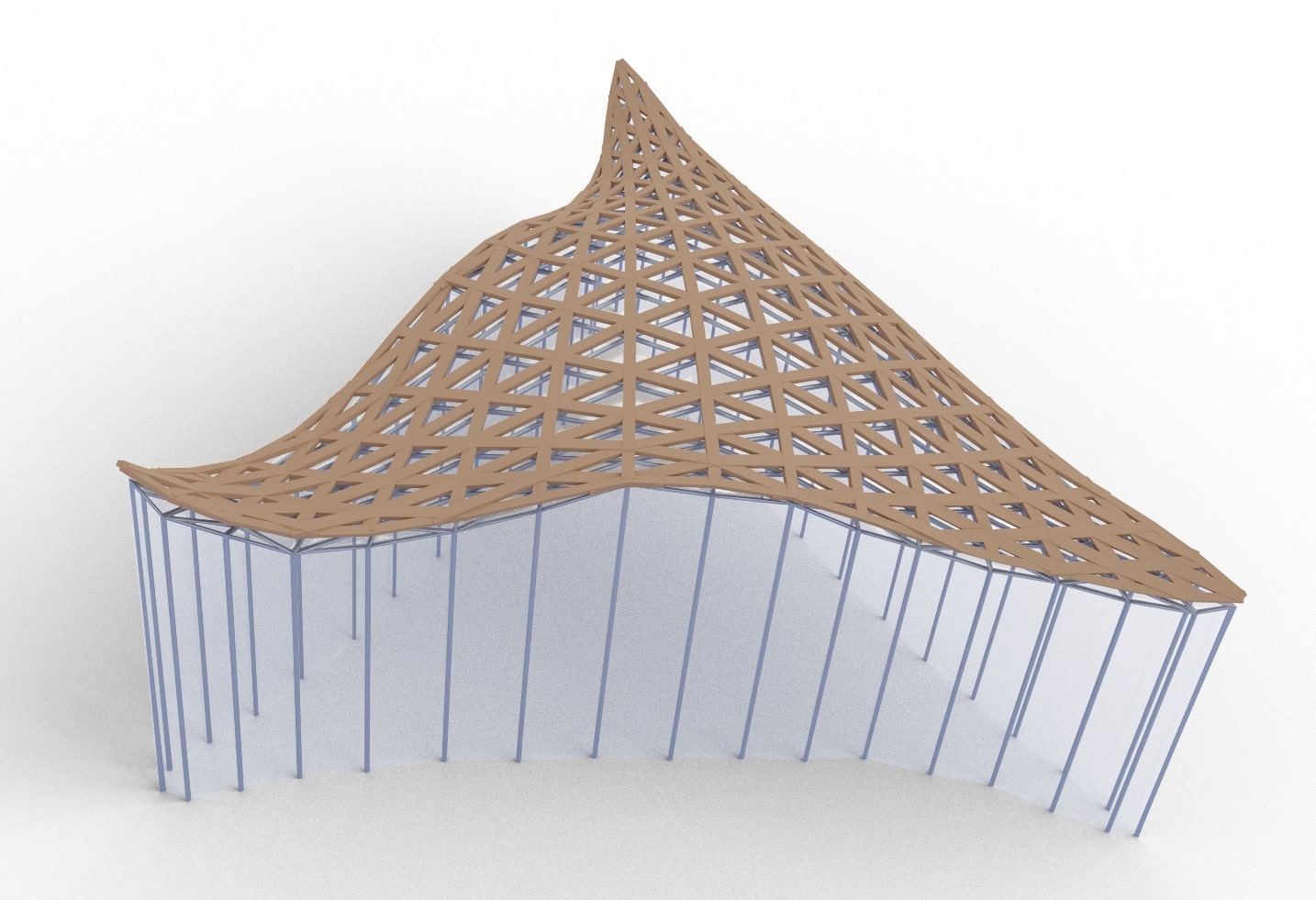}
\small
\put(0,58){\contour{white}{(b)}}
 \end{overpic}
\caption{(a) The gridshell from Figure~\ref{fig:CRPC2}(c) from a different point of view. (b) Another gridshell example extracted from a Euclidean GGG web obtained by optimizing an isotropic one.}
\label{fig:GGG}
\end{figure}
A direct computation shows that~\eqref{eq-taylor} satisfies~\eqref{eq-complex-crpc} up to 
$O(\varepsilon^3)$, if 
\begin{equation}\label{eq-f2}
f^{(2)}(w)=2\mathrm{Re} \left(h(w)^2\right) \log|h'(w)|
\end{equation}
and~\eqref{eq-f0}--\eqref{eq-f1} hold. 
Eqs.~\eqref{eq-f0}--\eqref{eq-f2} are \emph{not} the most general form of solutions, but they give much freedom for the design.

For design, it is convenient to prescribe a polynomial $h'(w)$ first, then compute $h(w)$ and $g(w)$ by integration, and use~\eqref{eq-f0}--\eqref{eq-f2} to get $$f^{\mathrm{appr}}(w)=f^{(0)}(w)+f^{(1)}(w)\varepsilon +f^{(2)}(w)\varepsilon^2 /2.$$ The roots of $h'(w)$ correspond to the roots of $g''(w)$, $K$, and $H$, hence to \emph{flat points}, where more than two asymptotic lines can meet. At these points, the angle between the asymptotic curves can be different from $\gamma$, and we need to regularize $f^{(2)}(w)$, e.g., 
replacing $\log|h'(w)|$ with $\log(|h'(w)|+\varepsilon)$ in~\eqref{eq-f2}. Then $z=f^{\mathrm{appr}}(w)$ is the desired approximate isotropic \bluer{CRPC} surface.

For instance, Algorithms~\ref{alg:C2L} and~\ref{alg:boundary+features} construct an approximate isotropic \bluer{CRPC} surface with given flat points or/and boundary. Optimization leads to Euclidean \bluer{CRPC} surfaces. 

We emphasize that we can control (some of) the flat points of the surface (see Figs. ~\ref{fig:CRPC3}, \ref{fig:CRPC2},  and~\ref{fig:GGG}(a)) and not just the boundary, 
as in the known design method for Weingarten surfaces by \cite{Pellis2021}. 

This section demonstrates one more advantage of isotropic geometry: it often comes with a powerful toolbox of complex analysis.

\section{Optimization\label{sec:optimization}}
We transform an isotropic shape into a Euclidean one \bluer{by optimization. 

For quad-mesh mechanisms, we apply the optimization algorithm from our previous work~\cite{quadmech-2024} as a black box, initializing it with isotropic mechanisms constructed in Sec.~\ref{sec:flexible}. The idea of the algorithm is to achieve easily-characterized \emph{infinitesimal flexibility} and \emph{sufficiently many isometric mechanism positions} (about ten); then a general theorem guarantees flexibility. We refer to a detailed description of the algorithm in~\cite{quadmech-2024} and focus on webs and CRPC surfaces in this section. 

For webs and CRPC surfaces,} we use the \emph{guided projection algorithm} \cite{Tang2014}. This algorithm is a regularized Gauss-Newton method, also known as Levenberg-Marquardt method. 
A key idea is to enhance performance by using at most quadratic constraints. To achieve that, auxiliary variables are introduced.







Also, we introduce a \emph{special inner product} of two vectors \( p = (p_1, p_2, p_3) \) and \( q = (q_1, q_2, q_3) \) as follows:
\begin{equation}
\langle p, q \rangle_\varepsilon = p_1q_1 + p_2q_2 + \varepsilon p_3q_3,
\label{eq:epsilon-metric}
\end{equation}
where \( \varepsilon = 0 \) represents the isotropic metric, and \( \varepsilon = 1 \) corresponds to the Euclidean metric. During the optimization process, \( \varepsilon \) is gradually increased from \( 0 \) to \( 1 \) to ensure a smooth transformation and improve the convergence behavior.

\subsection{Constraints}


We introduce the following auxiliary variables and constraints.

\textbf{Discrete A-nets.}
Consider an A-net $f_{ij}$, so that 
$f_{ij}, f_{i-1, j}, f_{i, j-1}$, $f_{i+1, j},$ and $f_{i, j+1}$ are coplanar for each $0<i,j<n$. This condition is encoded by the quadratic constraints
\begin{equation}\label{const:asym1}
     \langle \mathbf{n}_{ij}, f_{ij} - f_{i\pm 1, j} \rangle = 0, \quad \langle\mathbf{n}_{ij},  f_{ij} - f_{i, j\pm 1}\rangle = 0,  \quad \|\mathbf{n}_{ij}\|^2 - 1 = 0,
\end{equation}
where $\mathbf{n}_{ij}$ is an auxiliary variable representing the Euclidean normal to the net at $f_{ij}$, and $\langle a,b \rangle$ and $\|a\|$ are the Euclidean inner product and norm. Here the normalization of $\mathbf{n}_{ij}$ is essential to avoid convergence of \( \mathbf{n}_{ij} \) to $0$ during optimization.

The objective energy $E_A$ is the sum of the squared left sides of~\eqref{const:asym1} 
over all $0<i,j<n$. If the $(i-j)$- and $(i+j)$-lines  of $f_{ij}$ form two A-nets, the energy is defined analogously. 

\textbf{Discrete geodesics.}
Consider a discrete curve \( C \) with the vertices \( f_0, f_1, \dots, f_k \) on a discrete surface.  
Recall that the curve \( C \) is a discrete geodesic if
the surface normal is orthogonal to the curve binormal at each vertex $f_1, \cdots, f_{k-1}$.  
Orthogonality is expressed using the inner product, but instead of the Euclidean one, we employ the special inner product \(\langle \cdot, \cdot \rangle_\varepsilon\).  
This makes the geodesic constraints quadratic:
\begin{equation}\label{constr:binormal1} 
\langle \mathbf{b}_{i}, f_{i} - f_{i\pm 1} \rangle_\varepsilon = 0,  \thinspace \thinspace \langle \mathbf{b}_{i}, \mathbf{n}_{i} \rangle_\varepsilon = 0, \thinspace \thinspace \|\mathbf{b}_i\|^2 - 1 = 0, \thinspace \thinspace \|\mathbf{n}_i\|^2 - 1 = 0,
\end{equation} 
where \( \mathbf{b}_i \) is an auxiliary variable representing the binormal of \( C \) at \( f_i \) and 
\( \mathbf{n}_i \) is an auxiliary variable representing the normal of the discrete surface at \( f_{i} \) (we take a single variable \( \mathbf{n}_i \) per surface point, even if several curves $C$ pass through the same point).
The normalization of \( \mathbf{b}_i \) and \( \mathbf{n}_i \) follows the same reasoning as for \( \mathbf{n}_{ij} \) above.


The energy \( E_G^{C,\varepsilon} \) is the sum of the squared left sides of~\eqref{constr:binormal1} over all $0 <i < k$.

\textbf{Fairness.}
To enforce smoothness of $C$, we apply a soft fairness constraint. If \( f_i \) is not connected to the boundary, we require $f_i$ to
be close to the midpoint of \( f_{i-1} f_{i+1} \):  
\begin{equation}\label{constr:fairness}  
2f_i - f_{i-1} - f_{i+1} \approx 0.  
\end{equation}  
Otherwise, \( f_{i-1}, f_i, \) and \( f_{i+1} \) are required to be close to collinear:  
\begin{equation}\label{constr:fairness1}  
\frac{f_i - f_{i-1}}{\|f_i - f_{i-1}\|} - \frac{f_{i+1} - f_{i}}{\|f_{i+1} - f_{i}\|} \approx 0.  
\end{equation}  
The fairness energy \( E^{C}_f \) is the sum of the squared left sides of either~\eqref{constr:fairness} or~\eqref{constr:fairness1} over all $0< i <k$.

\textbf{Closeness to the previous-iteration vertices.} 
Consider selected vertices $f_0, f_1, \cdots f_s$ of $f_{ij}$.  A self-closeness constraint on $f_i$ ensures that 
 each vertex 
\( f_{i} \) remains close to its previous position \( f^{pr}_{i} \) for $0\le i \le s$:  
\begin{equation}\label{constr:vertices}  
\| f_{i} - f^{pr}_{i} \| \approx 0.
\end{equation}  
The vertex control energy \( E_{vcl} \) is the sum of the squared left side of~\eqref{constr:vertices} over all $0 \le i\le s$.  
The role of $E_{vcl}$ is to keep the selected vertices close to their previous positions, stabilizing the iterative procedure and preventing large updates. We use the squared norm to keep the term at most quadratic, as required by our framework.

\textbf{Closeness to a reference curve.}
If \( C \) approximates a fixed smooth curve, the vertices of \( C \) can move while preserving the overall shape of the curve. We enforce closeness by restricting each vertex \( f_i \) to move along directions close to the unit tangent \( \mathbf{e}_1 \) of the curve at its closest point \( f^{cl}_{i} \) on the curve:
\begin{equation}\label{curve:close}  
\langle f_{i} - f^{cl}_{i},  \mathbf{e}_{2}\rangle \approx 0, \quad  
\langle f_{i} - f^{cl}_{i},  \mathbf{e}_{3} \rangle \approx 0,  
\end{equation}  
where \( \mathbf{e}_1, \mathbf{e}_2, \mathbf{e}_3 \) form an orthonormal basis.  
The closeness energy $E^C_{ccl}$ is the sum of the squared left side of~\eqref{curve:close} over all $0\le i \le k$.

\textbf{Closeness to the previous-iteration surface.} 
For better convergence, the vertices should stay 
close to the previous iteration surface. 
This is enforced 
by requiring $f_{ij}$  to move 
close to the tangent plane 
at its previous position, $f^{pr}_{ij}$: 
\begin{equation}\label{close} 
\langle f_{ij} - f^{pr}_{ij}, \mathbf{n}^{pr}_{ij} \rangle \approx 0, 
\end{equation} 
where $\mathbf{n}^{pr}_{ij}$ is the normal of the previous iteration surface at $f^{pr}_{ij}$. The closeness energy $E_{scl}$ is the sum of the squared left side of~\eqref{close} for $0<i,j<n$.

\textbf{Constant angle.}
To measure the intersection angle of parameter lines at \( f_{ij} \), we use the angle between the central lines 
of the face \( f_{ij}f_{i+1,j}f_{i+1,j+1}f_{i,j+1} \) \cite{CRPC-2021}. Let \( f_a, f_b, f_c, f_d \) be the midpoints of its edges. The angle constraint, ensuring a constant angle $\gamma$ between central lines, is

\begin{equation}
\left\langle \frac{f_c - f_a}{\|f^{pr}_c - f^{pr}_a\|}, \frac{f_d - f_b}{\|f^{pr}_d - f^{pr}_b\|} \right \rangle -\cos \gamma = 0.
\label{eq:angle_constraint}
\end{equation}
Angle constraints ~\eqref{eq:angle_constraint} are quadratic as \(\|f^{pr}_c - f^{pr}_a\|\) and \(\|f^{pr}_d - f^{pr}_b\|\) are taken from the previous iteration. The angle energy \( E_{angle} \) is the sum of the squared left sides of ~\eqref{eq:angle_constraint} for all faces that do not contain flat points or boundary vertices.
\subsection{Energies}
\textbf{GGG Webs.} 
The objective $E_{GGG}^{\varepsilon}$ for a GGG web  for a given \( \varepsilon \)  is
the sum of energies \( E_{G}^{C, \varepsilon} \) of all non-boundary $i$-, $j$-, and $(i-j)$-lines respectively. 



\textbf{AAG webs.} 
The objective $E_{AAG}^{\varepsilon}$  for an AAG web for a given \( \varepsilon \) is 
the sum of  $E_A$ and the energies \( E_{G}^{C, \varepsilon} \) of all non-boundary $(i-j)$-lines.





\textbf{AGAG webs.} 
The objective for an AGAG web for a given \( \varepsilon \) is:  
\begin{equation}\label{energy:agag}  
E_{AGAG}^{\varepsilon} = E_{G}^{i, \varepsilon} + E_{G}^{j, \varepsilon}  + E^{even}_{A} + E^{odd}_{A},  
\end{equation}  
where \( E_{G}^{i, \varepsilon} \) and \( E_{G}^{j, \varepsilon} \) are the sums of energies \( E_{G}^{C, \varepsilon} \) of all non-boundary $i$- and $j$-lines respectively, and
 \( E^{even}_{A} \) and \( E^{odd}_{A} \) are the energies $E_A$ of the two  A-nets formed by $(i-j)$- and $(i+j)$-lines. 




\textbf{CRPC Surfaces.} 
The objective for a CRPC surface is  
\begin{equation}\label{energy:crpc}  
E_{CRPC} = E_{A} + E_{angle} +  E^{\partial}_{ccl} + E^{\partial}_{vcl}.
\end{equation}  
If the surface boundary is prescribed, \( E^{\partial}_{ccl} \) is the closeness energy for the boundary curve, and \( E^{\partial}_{vcl} \) is the closeness vertices energy for the boundary vertices. Otherwise, these two terms are omitted.

\textbf{Main Objective.} 
The non-linear least-squares problem for a given \(\varepsilon\) aims to minimize  
\begin{equation}\label{eq:epsenergy}
E^{\varepsilon} = E_{hard} + \omega_{0}E_{f} + \omega_{1}E_{scl} + \omega_{2}E_{vcl},
\end{equation}  
where \(E_{hard} \in \{E^{\varepsilon}_{GGG}, E^{\varepsilon}_{AGAG}, E^{\varepsilon}_{AAG}, E_{CRPC}\}\) represents hard constraints. The energy \(E_{f}\) ensures smoothness, while $E_{scl}$ and $E_{vcl}$ prevent drastic shape changes during the optimization (here $E_{vcl}$ is the sum over all vertices $f_{ij}$).
The weights \(\omega_{0}, \omega_1\), and \(\omega_{2}\) balance fairness and approximation.

\subsection{Initialization}\label{init}  
For webs and mechanisms, the shapes are initialized using their isotropic counterparts 
constructed in Sections~\ref{sec:flexible} and~\ref{sec:webs}. For CRPC surfaces, initialization is based on the second-order approximation of the isotropic CRPC surface constructed using either Algorithm~\ref{alg:C2L} or~\ref{alg:boundary+features}. The resulting surface is then remeshed using the libigl~\cite{libigl} implementation of mixed-integer quadrangulation (MIQ)~\cite{quadrangulation}.
Now we provide initialization details for the auxiliary variables.

\textbf{GGG webs.} 
The binormals $\mathbf{b}_i$ and the normals $\mathbf{n}_i$ are initialized as the discrete binormals and $(0,0,1)$, respectively. 

\textbf{AGAG and AAG webs.} 
Two sets of auxiliary normals are introduced: one for the A-net constraints, initialized using discrete Euclidean normals, and another for the geodesic constraints. The latter, along with the binormals, are initialized as in the GGG case. At \(\varepsilon = 0\), all constraints are satisfied since the A-nets are the same in both geometries.



\textbf{CRPC surfaces.}  
The normals $\mathbf{n}_{ij}$ are initialized as \bluer{the normalized cross product $(f_{i+1,j} - f_{i-1,j}) \times (f_{i,j+1} - f_{i,j-1})$}.


\begin{figure}[t]
    \centering
 \begin{overpic}
[width=1.0\linewidth]{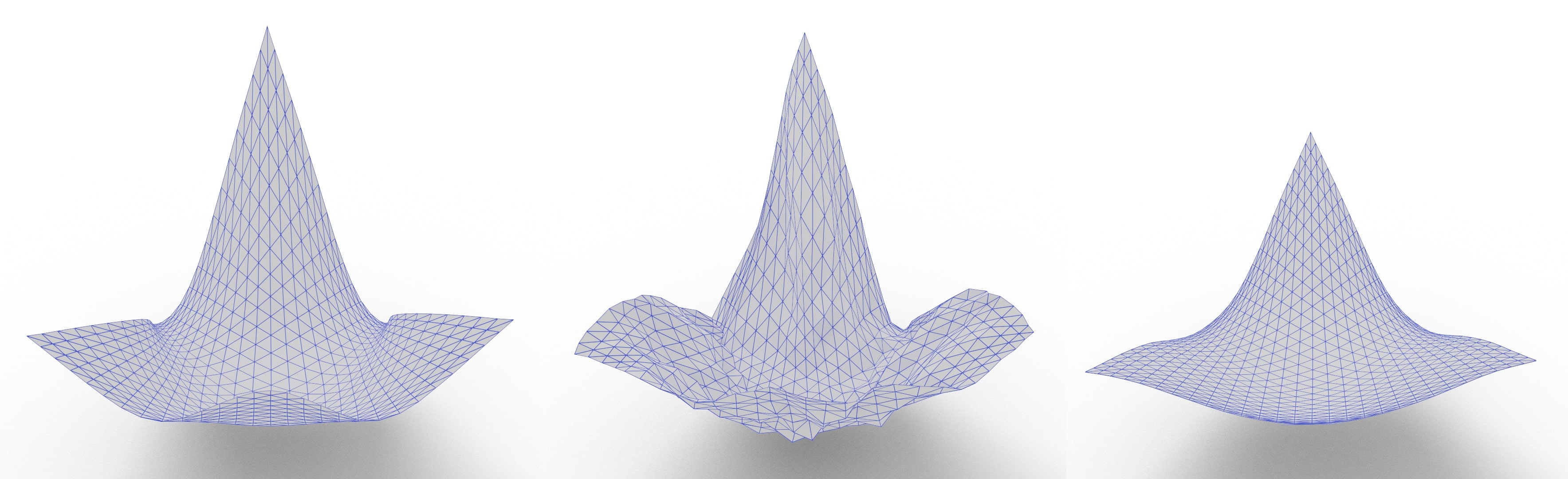}
\small
\put(0,27){\contour{white}{(a)}}
\put(32,27){\contour{white}{(b)}}
\put(65,27){\contour{white}{(c)}}
 \end{overpic}
\caption{Advantage of using the gradual approach with metric~\eqref{eq:epsilon-metric}. We compare Euclidean GGG webs obtained by optimizing the initial isotropic GGG web (Fig.\ref{fig:GGG_merged}(e)) under different conditions. (a) The web optimized with the gradual approach and the set of weights from Table~\ref{tab:results}. (b) The web optimized without the gradual approach, using the same set of weights; the shape becomes non-smooth. (c) The web optimized without the gradual approach but with a higher fairness weight; here, the shape becomes significantly flatter.
}
\label{fig:GGG_fail}
\end{figure}

\subsection{Optimization Process} 
We minimize the energy \( E^{\varepsilon} \) given by~\eqref{eq:epsenergy} iteratively for \(\varepsilon\) increasing from $0$ to $1$, using the result of each step as the initialization for the next. At each step, we ensure that the hard constraints \( E_{hard} \) reach the accuracy of \(10^{-5} \). Table~\ref{tab:results} 
summarizes the choices of weights, accuracy, and the number of variables for the final iteration with \(\varepsilon = 1\). For the other $\varepsilon$, 
the weights are the same. 

The usage of the gradual approach with  metric~\eqref{eq:epsilon-metric} plays a critical role (see Fig.~\ref{fig:GGG_fail}). For less constrained webs (i.e., GGG and AAG webs), it prevents drastic changes from the initial shapes. In the case of AGAG-webs its role is even more significant, as optimization consistently failed without this approach (see supplementary materials for details).

For quad-mesh mechanisms, the initial guesses via isotropic mechanisms
work extremely well with the optimization algorithm of
\cite{quadmech-2024}; 10 iterations have been sufficient to satisfy the constraints
with high accuracy without using the gradual approach based on 
metric (\ref{eq:epsilon-metric}). Supplementary materials include a video showing the isotropic and Euclidean flexion of Figs.~\ref{fig:flexible-isotropic-i},~\ref{fig:flexible-isotropic-ii}, and~\ref{fig:flexible-isotropic-projective} (Euclidean only).
\begin{table*}[t]
\centering
\begin{tabular}{|l|c|c|c|c|c|c|c|c|}
\hline
\textbf{\small Fig.} & \(\lvert V\rvert\) & \(N_v\) &\(bbd\) & \(\omega_1\) & \(\omega_2\) & \(\omega_3\) & \textbf{\small T/iter } & \(E_{hard}\) \\
\hline
\small ~\ref{fig:teaser}   & \small 1985  & \small 10912&\small 3.99 & \small 5e-3 & \small 5e-3 & \small 5e-3 & \small 0.74 s & \small 2.7e-12 \\
 \small ~\ref{fig:AGAG1}  & \small 361  & \small 4551 & \small 4.54& \small 1e-2 & \small 1e-2 &  \small 1e-2 & \small0.11 s & \small 1e-14 \\
 \small ~\ref{fig:CRPC1} & \small 873 & \small 4381& \small 4.45 & \small 5e-3 & \small 5e-3 & \small 5e-3 & \small0.4 s & \small 4.9e-12 \\
\small ~\ref{fig:GGG_merged} & \small 631  & \small 8457&\small 34.43, 292.98 & \small 1e-3 & \small 1e-3 & \small 1e-3 & \small0.15 s & \small 4.8e-20, 3.7e-20 \\
\small ~\ref{fig:AGAG_merged} & \small 1369  & \small 18807&\small 4.50, 1.53 &  \small 1e-2 & \small 1e-2 & \small 1e-2 & \small0.36 s & \small 2.7e-12, 3.1e-10 \\
\small ~\ref{fig:AAG1}  & \small 441  & \small 4572& \small 2012.80 & \small 5e-3 & \small 5e-3 & \small 5e-3 & \small0.13 s & \small 3.9e-20 \\
\small ~\ref{fig:AAG21} & \small 1681 & \small 18732& \small 3239.26 & \small 5e-3 & \small 5e-3 & \small 5e-3 & \small0.3 s & \small 8.9e-20 \\
\small ~\ref{fig:CRPC3} & \small 1309 & \small \small 7084& \small 6.71 & \small 5e-3 & \small 5e-3 & \small 5e-3 & \small0.6 s & \small 4.9e-12 \\
\small ~\ref{fig:CRPC2} & \small 757 & \small 3823 &\small 6.04&  \small 5e-3 & \small 5e-3 & \small 5e-3 & \small0.38 s & \small 4.9e-12\\
\small ~\ref{fig:GGG}(b)  & \small 154 & \small 1734 &\small 187.14& \small 1e-3 & \small 1e-3 & \small 1e-3 & \small0.08 s & \small 1.1e-20 \\
\hline
\end{tabular}
\caption{Optimization statistics for selected figures. \bluevar{\(\lvert V \rvert\) and \(bbd\) denote the number of vertices and the diagonal length of the axis-aligned bounding box of the discrete surface, respectively}.  \(N_v\) is the number of variables in the optimization process. 
the surface. $\omega_0$, $\omega_1$, and $\omega_2$ are the weights at the first iteration for the fairness term and for approximation terms for each $\varepsilon$. T/iter represents the average time per iteration. \(E_{hard}\) is the energy of the hard constraints, namely GGG, AGAG, AAG, and CRPC constraints at the last iteration.}
\label{tab:results}
\end{table*}

\section{Discussion and Conclusion\label{sec:conclusion}}
\textbf{Implementation.}
The optimization has been implemented in Python and tested on an Intel Xeon E5-2687W 3.0 GHz processor.
Typically, around 10–20 iterations are required to reach the convergence for each value of $\varepsilon$. The weights $\omega_1, \omega_2$, and $\omega_3$ are reduced by $10$ times every five iterations, with the final reduction setting them to zero to ensure strict convergence. In Table~\ref{tab:results}, we observe that the time per iteration for CRCP surfaces is higher than for the other webs. This increase is due to angle constraint~\eqref{eq:angle_constraint}, which requires computing the length of each edge and which is not needed for the other cases. 

\textbf{Advantages.} Our approach has advantages compared to known ones.
The GGG webs appearing in architecture are mostly spherical patches. This 
matches the existence theorem for GGG webs on surfaces with constant Gaussian curvature by Mayrhofer~\cite{mayrhofer:1931}.
In recent work~\cite{web-propagation-2025}, GGG and AAG webs are constructed by propagation. There, they take the first strip, consisting of two curves, and propagate the web by adding curves one by one and re-optimizing the whole shape. The propagation algorithm can only produce regular grids without singularities. The shape control is biased: it is defined and controlled by the first strip. The areas far from it need extra manual control to create variation; otherwise, the surface eventually becomes flat except for the areas close to the initial strip. The present paper overcomes these limitations.

Schling et al.~\cite{Eike-CAD-2022} discuss AAG and AGAG webs. They initialize the optimization of AAG webs with negatively curved rotational surfaces, thus limiting the variety of shapes. Their examples of AGAG webs are mostly flat and symmetric, due to the lack of initialization. Our initialization provided much more freedom.
%

Previous constructions of CRPC surfaces 
use optimization keeping the 
topology of the surface. We are not aware of a single approach known before that allows control of the positions of flat points.





The only known large-sized quad mesh mechanisms are Voss nets and T-nets. In the work~\cite{quadmech-2024}, the authors discuss several ways to initialize and optimize general quad mesh mechanisms. \bluer{The mechanisms were constructed there by subdivision of small-sized meshes or known pairs of isometric surfaces. The former strategy has limited constructional degrees of freedom.
The latter requires remeshing, and the resulting initializations are often far from the true solutions, thereby exacerbating optimization difficulties and
increasing the risk of convergence failure.} The limitation in the variety of shapes again comes from initialization, which is resolved in the present work. \bluer{Moreover, as observed in Sec.~\ref{sec:flexible}, initialization with a single isotropic mechanism yields an entire family of Euclidean ones, depending on several parameters.}

\textbf{Limitation.} \bluer{The main limitation of the suggested approach is the necessity to find and solve a nondegenerate analog of the original Euclidean problem in isotropic geometry.}
For the CRPC surfaces, our initialization is \bluer{additionally} limited to flat points where four asymptotic lines meet, and our remeshing step uses a cross-field-based algorithm, which limits the placement of flat points. In some cases, flat points may be split into multiple nearby ones to satisfy the quality requirements of the remeshing tool (see Fig.~\ref{fig:CRPC2}). Developing new techniques for more accurate control over flat points placement and remeshing remains an important direction for future work.

\begin{figure}[htbp]
    \centering
 \begin{overpic}
[width=0.85\linewidth]{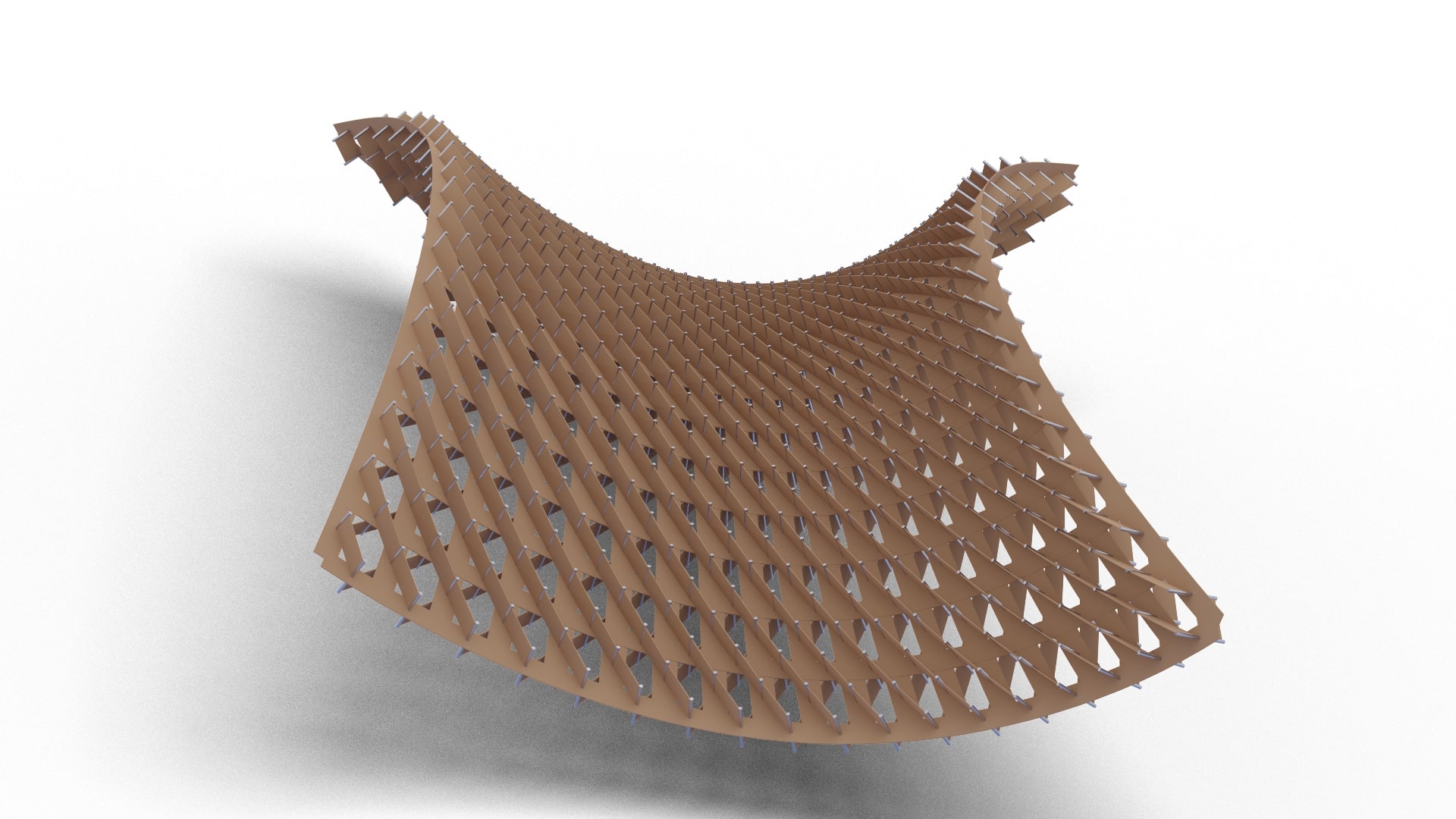}
\small
 \end{overpic}
\caption{The gridshell from Figure~\ref{fig:AGAG_merged}(h) from a different point of view.} 
\label{fig:AGAG2grid}
\end{figure}

\textbf{Shape change.} For the CRPC surfaces with given flat points, GGG-  and AAG-webs, and flexible nets, the visual change after optimization from isotropic to Euclidean geometry is minimal 
(see Figs.~\ref{fig:flexible-isotropic-i},\ref{fig:GGG_merged},\ref{fig:AAG1},\ref{fig:AAG21},\ref{fig:CRPC2},  and~\ref{fig:GGG}). In contrast, CRPC surfaces with prescribed flat points and boundary and AGAG-webs, the shape change after optimization can be drastic  (see Figs.~\ref{fig:AGAG_merged}(g), \ref{fig:AGAG2grid},  and~\ref{fig:CRPC1}). The latter cases are also the hardest to optimize due to being geometrically overconstrained. 

\textbf{Conclusion.} 
In this work, we demonstrated the efficiency of iso\-tropic geometry as a powerful framework for initializing and solving Euclidean geometric design problems by optimization. Our contributions include methods to construct 
approximate isotropic CRPC surfaces and 
discrete isotropic AAG webs, which serve as effective initializations for their Euclidean counterparts.
We also studied discrete AGAG webs and quad mesh mechanisms with planar faces, where the recent classification of their isotropic versions are used as initial data for optimization.

\textbf{Future Work.}
A natural direction for future research is the question of whether one can smoothly transform a minimal surface (iso\-tropic or Euclidean) with a given boundary into a CRPC surface (isotropic or Euclidean) with the same boundary. In a forthcoming publication \cite{SkopenkovYorov2025CRPC}, we prove that such a transformation is possible in the absence of flat points. The case of surfaces with flat points remains open.

Another direction is to explore webs composed of circular flat lamellas instead of straight ones. The case where the lamellas are orthogonal to the reference surface and follow its parameter lines (known as \emph{S-nets}) has been studied in~\cite{Pellis2020}. S-nets are characterized by both families of parameter lines having constant normal curvature. An interesting direction is to consider circular lamellas that are tangent to the reference surface. These lamellas follow curves of constant geodesic curvature. In the isotropic setting, the top view of such a web consists of circular arcs, forming a planar web composed of such arcs. While the planar case of circular webs remains an open problem, several classification results are available (see~\cite{MR610443, pottmann2012darboux, hexagonalwebs}).

Another direction for further applications of the proposed method is the structural design with non-vertical loads. In particular, one may extend the analysis by \cite{Amendola-etal-22} of forces that can be supported by a given 2D structure to the 3D case.



\section*{Acknowledgement}
\vspace{-0.1cm}
This work was supported in part by the Beijing Natural Science Foundation (Z240002), the Key Program of the National Natural Science Foundation of China (12494550, 12494553), the National Natural Science Foundation of China (62495092, 62125305), the NSFC–FDCT Joint Project (62461160259), and KAUST Baseline Funding.

\bibliographystyle{elsarticle-num}
\bibliography{references.bib}


\end{document}